\documentclass[10pt,aps,twocolumn,pra,superscriptaddress,nofootinbib]{revtex4-2}

\usepackage[T1]{fontenc}
\usepackage{dsfont}
\usepackage{amsfonts}
\usepackage{amsmath}
\usepackage{bm}
\usepackage{amssymb}
\usepackage{graphicx}
\usepackage{braket}
\usepackage{mathtools}

\usepackage{soul}

\usepackage[colorlinks=true,urlcolor=blue,citecolor=blue,linkcolor=blue]{hyperref}

\newcommand{\al}{\alpha}

\newcommand{\la}{\lambda}

\newcommand{\si}{\sigma}

\newcommand{\La}{\Lambda}

\newcommand{\Nega}{\mathcal{N}}    
\newcommand{\diff}{\mathop{}\!\mathrm{d}}
\newcommand{\OQRI}{\mathrm{OQR}_I}
\newcommand{\OQRInv}{\mathrm{OQR}_{I^{-1}}}

\newcommand{\Rr}{\mathsf{R}}
\newcommand{\Hs}{\mathcal{H}}
\newcommand{\HSs}{\rema{\mathcal{B}}}

\newcommand{\IMA}{\textrm{im}}

\def \Tr {\text{Tr}}
\newtheorem{theorem}{Theorem}
\newtheorem{lemma}{Lemma}

\makeatletter
\g@addto@macro\bfseries{\boldmath}
\makeatother

\newcommand{\rema}[1]{\textcolor{black}{#1}}

\begin{document} 

\title{Quantum metrology of rotations with mixed spin states}
\author{Eduardo Serrano-Ens\'astiga}
\email{ed.ensastiga@uliege.be}
\affiliation{Institut de Physique Nucléaire, Atomique et de Spectroscopie, CESAM, University of Liège,
B-4000 Liège, Belgium}

\author{Chryssomalis Chryssomalakos}
\email{chryss@nucleares.unam.mx}
\affiliation{Instituto de Ciencias Nucleares,  Universidad Nacional Autónoma de México PO Box 70-543, 04510, CDMX, México}
\author{John Martin}
\email{jmartin@uliege.be}
\affiliation{Institut de Physique Nucléaire, Atomique et de Spectroscopie, CESAM, University of Liège,
B-4000 Liège, Belgium}
%
%
\begin{abstract}
\rema{The efficiency of a quantum metrology protocol can be significantly diminished by the interaction of the system with its environment, leading to a loss of purity and, as a result, a mixed state for the probing system. An example is the measurement of a magnetic field through the rotation of a spin that is subject to decoherence due to its coupling to a surrounding 
spin or bosonic bath. In this work, we define mixed optimal quantum rotosensors (OQRs) as mixed spin-$j$ states that achieve maximum sensitivity to estimate infinitesimal rotations, when the rotation axis is unknown. We study two scenarios, where the probe states saturate the averaged fidelity or the averaged quantum Cramér-Rao bound, the latter giving the ultimate sensitivity. We find that mixed OQRs can achieve sensitivity equal to that of pure states and are obtained by mixing states from linear subspaces of anticoherent states. We present several examples of mixed OQRs and their associated anticoherent subspaces. We also show that OQRs maximize entanglement in a specific sense, preserving the known relation between entanglement and optimal rotation sensitivity for pure states, even in the context of mixed states. Our results highlight the interconnection between quantum metrology of rotations, anticoherence and entanglement in mixed spin states.} 
\end{abstract}
\maketitle
\section{Introduction}
The development of modern measurement techniques is closely tied to advances in all fields of science. The unprecedented manipulation and control of quantum systems is now opening up opportunities for quantum-enhanced metrology, which has become a major field of study. A fundamental example is the precise detection of rotation, first developed for inertial navigation and now used to detect minute magnetic fields through the spin of particles. The study of the measurement of a rotation using a quantum system as a probe reveals the existence of certain states where the estimate can be improved by quantum effects~\cite{science.1104149,RevModPhys.89.035002,RevModPhys.90.035005}. The main tool for estimating the sensitivity of the state of a quantum system is the quantum Cramér-Rao bound (QCRB)~\cite{Bra.Cav:94,helstrom1969quantum}. 
\rema{When the axis of rotation is distributed randomly and uniformly over the sphere, the optimal states are those that maximize sensitivity averaged over all rotation axes.} 
 This corresponds exactly to the notion of an optimal quantum rotosensor (OQR), which has been extensively studied for pure spin-$j$ states in recent years~\cite{Kolenderski_Demkowicz-Dobrzanski_2008,Chr.Her:17,Gol.Jam:18,Mar.Wei.Gir:20,ZGoldberg_2021,Bou.etal:17}. In particular, it was discovered that the OQRs for infinitesimal rotations are anticoherent (AC), a property of quantum states first introduced by Zimba~\cite{Zim:06} and studied in many works, see, e.g., \cite{Gir.Bra.Bag.Bas.Mar:15,Bag.Bas.Mar:14,Bag.Dam.Gir.Mar:15,Bag.Mar:17,Rudzinski2024orthonormalbasesof,Crann_2010,Björk_2015,Den.Mar:22}. An anticoherent state is characterized by \rema{vanishing of the angular momentum expectation value~\cite{Zim:06,Gir.Bra.Bag.Bas.Mar:15}, that is, they are states that ``point nowhere''}. In this sense, they are the opposite to spin-coherent states~\cite{Rad:71,Zim:06}.

Under real conditions, a quantum system interacts with its environment and its state becomes mixed, i.e., a statistical mixture of pure states. \rema{Several scenarios encompass this situation such as: errors in the preparation of a pure state, loss of particles from the
probe system, decoherence produced by the interaction with the environment, or systematic errors~\cite{PhysRevA.81.022106,PhysRevA.82.042114,PhysRevX.1.021022,Den.Mar:22,Pedram_2024}, to mention a few. In this framework}, it is not clear that the conditions required for a mixed state to be an OQR are the same as those defining pure OQRs. 
While the characterization of the most sensitive fixed-spectrum mixed states for an infinitesimal rotation over a known axis, and in general for an infinitesimal transformation given by any generator of a Lie algebra, has been solved in Ref.~\cite{PhysRevLett.123.250502}, it remains open for an unknown axis, \rema{i.e., in the above, average sense}. Additionally, the authors of Ref.~\cite{PhysRevA.92.032317} have characterized the mixed states with the best metrological power via the purification of the state. In the present work, precisely, we derive conditions for obtaining the maximal detection sensitivity for an infinitesimal rotation about an unknown axis with mixed states. We show how to write these conditions in terms of anticoherent subspaces, defined as vector spaces of anticoherent states, which were first introduced and studied by Pereira and Paul-Paddock~\cite{Per.Pau:17}. It turns out that these AC subspaces also have applications in a holonomic quantum computation approach resilient to rotational noise~\cite{Toponomic:22}.

\rema{It is well established that} multipartite entanglement \rema{can be considered as}  a resource to achieve the ultimate sensitivity of measurement with quantum states~\cite{doi:10.1142/S0219749909004839,RevModPhys.90.035005,giovannetti2011advances,PhysRevB.107.035123,PhysRevLett.102.100401,PhysRevA.85.022321,PhysRevA.85.022322,PhysRevA.82.012337}, known as the Heisenberg limit. \rema{In this paper, we also explore possible connections between entanglement and the metrological power of mixed states. More specifically, we study the relation between entanglement negativity~\cite{PhysRevA.58.883,PhysRevA.65.032314} and anticoherence.}
For pure states, \rema{negativity is}
a faithful measure of bipartite entanglement~\cite{PhysRevA.65.032314,Ben.Zyc:17}, \rema{and is maximized for certain AC states~\cite{Den.Mar:22}. For mixed states, although negativity is in general only a witness of entanglement~\cite{PhysRevLett.77.1413,HORODECKI19961}, we show a similar connection between entanglement and anticoherent mixed states.
For an individual spin-$j$, the notion of entanglement is a priori meaningless. It becomes meaningful when the spin $j$ is seen as the result of the addition of $N=2j$ spin-$1/2$ in a symmetric state. This picture provides insight into other non-classical properties of spin states, such as spin squeezing, which serves as a reference for metrological gains; see, for example, Refs.~\cite{PhysRevA.47.5138,Vitagliano2014,Fadel2020} for a discussion of this correspondence.}

The paper is organized as follows. After reviewing the basic concepts in Sec.~\ref{Sec.Concepts}, we derive the conditions for pure and mixed states to be OQR in Sec.~\ref{Sec.OQR}. We study the question of the existence of mixed OQRs for small spin and present illustrative examples in Secs.~\ref{Sec.OQR.small.spin}--\ref{Sec.Examples.OQR}, respectively. The discussion of the negativity of mixed OQR states is presented in Sec.~\ref{Sec.Neg.AC}. Finally, we summarize our results in Sec.~\ref{Sec.Conclusions}. A series of appendices detail the more technical results.
\section{Concepts and mathematical framework}
\label{Sec.Concepts}
\subsection{\rema{Fidelity, Quantum Fisher Information\\ and QCRB}}
\rema{Let us first consider a quantum system with a Hilbert space $\Hs^{(j)}$, acted upon  by the $(2j+1)$-dimensional (spin-$j$) $\mathrm{SU}(2)$ irreducible representation (irrep) --- 
more general systems will be considered in Subsection~\ref{SubSec.OQR.general}.
The fidelity between a state $\ket{\Psi} \in \Hs^{(j)}$ and its image $\ket{\Psi_\Rr}$ under the rotation $\Rr^{(j)} ( \mathbf{n}, \eta )$ is defined by
\begin{equation}
\label{Fidelity.pure.general}
 F \big( |\Psi\rangle, |\Psi_\Rr\rangle \big) \equiv \big|\langle\Psi| \Rr^{(j)} ( \mathbf{n}, \eta ) |\Psi\rangle \big|^2 \, ,
\end{equation}
with $\Rr^{(j)} (\mathbf{n} , \eta  ) = e^{-i\eta \,  \mathbf{J}\boldsymbol{\cdot} \mathbf{n} } $ and where the unit vector $\mathbf{n} = (n_x , n_y , n_z) $ and the angle $\eta \in [0,2\pi[$ are the axis-angle parameters, and $\mathbf{J} = (J_x , J_y , J_z)$ is the vector of $\mathfrak{su}(2)$ generators in the spin-$j$ irrep.}
Our main goal is to find the states that minimize \rema{this fidelity}
averaged over all possible axes of rotation, i.e., the states for which the quantity 
\begin{equation}
\overline{F} (\eta, \ket{\Psi}) \equiv
    \frac{1}{4\pi} \int 
    F \Big(\ket{\Psi} , \Rr^{(j)} (\mathbf{n} ,\eta ) \ket{\Psi} \Big) \,
    \diff \mathbf{n}  ,
\end{equation}
is minimal, where $\eta \ll 1$. For mixed states,  \rema{$\rho$ and $\sigma$}, in the Hilbert-Schmidt space of operators $\HSs (\Hs^{(j)})$, we use the Uhlmann-Jozsa fidelity $F(\rho, \sigma)$~\cite{UHLMANN1976273,doi:10.1080/09500349414552171}, which can be related to the Bures distance~\cite{Bra.Cav:94}
\rema{
\begin{equation}
\label{Eq.Fidelity.mixed}
    F(\rho , \si ) = 
    \left( \Tr \sqrt{\sqrt{\rho}\, \si  \sqrt{\rho}} \,
    \right)^2 .
\end{equation} }
Like any well-defined fidelity function for mixed states~\cite{Liang_2019}, the Uhlmann-Jozsa fidelity coincides with the fidelity~\eqref{Fidelity.pure.general} when the states are pure. 

\rema{The fidelity $F(\rho, \rho_{\Rr})$ between a state $\rho$ and the infinitesimally close rotated state $\rho_{\Rr}= \Rr^{(j)}(\mathbf{n},\eta)\,\rho\, \Rr^{(j)}(\mathbf{n},\eta)^{\dagger}$ can be approximated for an infinitesimal angle of rotation $\eta \ll 1$ by the Taylor expansion
\begin{equation}
\label{Eq.Fidelity.small}
 F(\rho, \rho_{\Rr}) = 1 - \frac{1}{4} I(\mathbf{n} , \rho ) \eta^2 + \mathcal{O} \left(\eta^3 \right) , 
\end{equation} 
with
\begin{equation}
\label{Def.QFI}
 I(\mathbf{n} , \rho )  \equiv \left. -2 \,\frac{\partial^2 F(\rho,\rho_{\Rr})}{\partial \eta^2} 
     \right|_{\eta=0} 
\end{equation}
the Quantum Fisher information (QFI)~\cite{Bra.Cav:94}. This quantity can also be understood as the distance, according to the Bures metric, between two infinitesimally close density matrices $\rho$ and $\rho_{\Rr}$ with $\eta \ll 1$~\footnote{The relationship $I = 4\diff_B$ between the QFI ($I$) and the Bures metric ($\diff_B$) has been discussed in several references~\cite{Bra.Cav:94,HUBNER1992239,Hans-JürgenSommers_2003}, including some discrepancies between these quantities when the rank of the density matrix changes~\cite{PhysRevA.95.052320}. An exact correspondence has been proven in Ref.~\cite{zhou2019}.}. Thus, the states that minimize the fidelity between themselves and their image under infinitesimal rotation are those that maximize the QFI. This quantity also appears in the QCRB~\cite{Bra.Cav:94,PhysRevX.1.021022,PhysRevLett.123.250502} that bounds the minimum variance $\text{Var}(\hat{\eta})$ of the estimator $\hat{\eta}$ of the parameter $\eta$,
\begin{equation}
\label{Eq.QCRB}
M \,\text{Var} [\hat{\eta}] \geqslant \frac{1}{I( \mathbf{n} , \rho )} ,    
\end{equation}
with $M$ the number of times the measurement is repeated.
}
\subsection{Multipole operators}
The multipole (or polarization) operators $T_{LM}$, with $L=0, \dots, 2j$ and $M=-L, \dots ,L$~\cite{Fan:53,Var.Mos.Khe:88,Agarwal_2012}, form an orthonormal basis with respect to the Hilbert-Schmidt inner product,
$ \Tr ( T_{L_1 M_1}^{\dagger} T_{L_2 M_2} ) = \delta_{L_1 L_2} \delta_{M_1 M_2}  $. Hence, a mixed state $\rho$ expanded in this basis reads
\begin{equation}
\label{rhoTLMexpansion}
    \rho = \sum_{L=0}^{2j} \sum_{M=-L}^L \rho_{LM} T_{LM} = \rho_0 + \sum_{L=1}^{2j} \sum_{M=-L}^L \rho_{LM} T_{LM}  \, , 
\end{equation}
where $\rho_{LM} = \Tr (\rho T^{\dagger}_{LM})$ and $\rho_0 = (2j+1)^{-1} \mathds{1}$ is the maximally mixed state. Since $T_{L M}^{\dagger} = (-1)^{M} T_{L ,-M}$~\cite{Var.Mos.Khe:88}, the hermiticity of the density matrix enforces that $\rho^{*}_{LM} = (-1)^M \rho_{L -M} $. Additionally, the $T_{LM}$'s are traceless operators except for $T_{00} = \mathds{1}/\sqrt{2j+1}$. The multipole operators can be constructed as products of the angular momentum operators $J_{\pm} = J_x \pm i J_y$ and $J_z$ (see, e.g., Eq.~(4), p.~44 of Ref.~\cite{Var.Mos.Khe:88}). For example, 
\begin{equation}
\label{Eq.multipole.and.angular.momentum}
\begin{aligned}
 &  T_{10} \propto J_z , \quad 
  T_{20} \propto \mathbf{J}^2 - 3 J_z^2 ,
\\
&    T_{1\pm 1} \propto J_{\pm} ,  \quad T_{2\pm 1} \propto J_z J_{\pm} + J_{\pm} J_z , \quad 
T_{2\pm 2} \propto J_{\pm}^2 .
\end{aligned}
\end{equation}
More details on multipole operators can be found in Refs.~\cite{Fan:53,Var.Mos.Khe:88,Agarwal_2012}. 
\subsection{Anticoherent states and subspaces}
\label{Subsec.anticoherence}
We begin with the definition of anticoherent states as given in Ref.~\cite{Zim:06}, extended here to mixed states.\\
\paragraph*{Definition.} A state $\rho \in \HSs (\Hs^{(j)})$ is \emph{$t$-anticoherent} if 
$ \Tr (\rho J_{\mathbf{n}}^K)$ is independent of $\mathbf{n}$ for $K=1 , 2, \dots, t$, where $J_{\mathbf{n}} \equiv \mathbf{n} \boldsymbol{\cdot}\mathbf{J}$ with $\mathbf{n}\in\mathbb{R}^3$ a unit vector. 
For the sake of brevity in this paper, we will use the abbreviation $t$-AC for $t$-anticoherent.\\

The condition of $t$-anticoherence is equivalent to \rema{the vanishing of} the expectation value of the multipole operators, i.e., \ $\langle T_{L M} \rangle = \Tr (\rho\,  T_{L M}) = 0$, for $L=1, \dots ,t$ and $M= -L , \dots , L$~\cite{Gir.Bra.Bag.Bas.Mar:15}. In particular, the 1-anticoherence condition for pure states reduces to $\bra{\psi} J_a \ket{\psi} = 0$ for $a=x,y,z$~\cite{Bag.Mar:17}. The $t$-anticoherence of a state $\rho$ can be quantified with the functions~\footnote{The functions~\eqref{Ant.Func} define measures of anticoherence for pure states, see Ref.~\cite{Bag.Mar:17} for more details.}
\begin{equation}
\label{Ant.Func}
    \mathcal{A}_t (\rho) = \frac{t+1}{t} \left[ 1- \Tr (\rho_t^2) \right]
\end{equation}
for integer \rema{$t$ and $1 \leqslant t\leqslant N-1$}, where $\rho_t= \Tr_{N-t}(\rho)$ is the reduced density matrix obtained after tracing out $N-t$ spin-$1/2$ among the $N$ making up the $N/2=j$ spin. These functions range from $0$, for spin-coherent states, to $1$, for $t$-AC states~\cite{Bag.Mar:17}. In addition, a state $\rho$ is $t$-AC if and only if $\mathcal{A}_K(\rho)=1$ for $K=1,2,\dots, t$~\cite{Gir.Bra.Bag.Bas.Mar:15}. By definition, $t$-anticoherence implies $t'$-anticoherence for $t' < t$. While for pure states there are no $t$-AC states for certain values of $t \leqslant 2j $~\cite{Bag.Mar:17}, one can always find, at least sufficiently close to the maximally mixed state, states which are AC to any order $t$ (see Appendix~\ref{App.t.anticoherence} for a proof). 

The notion of an anticoherent state was generalized in Ref.~\cite{Per.Pau:17} to subspaces as follows.\\[0pt]

\paragraph*{Definition.} A $k$-dimensional subspace $V$ of $\Hs^{(j)}$  
is $t$-AC if any state $\ket{\psi} \in V$ is $t$-AC. Such a subspace will be referred to as $(j,k,t)$ AC subspace.\\[0pt]

A necessary, but not sufficient, condition for a basis $B_V= \{\ket{\psi_m}\}_m$ to span a $t$-AC subspace $V$ is that its elements $\ket{\psi_m}$ be $t$-AC states. Orthonormal bases \rema{for the entire Hilbert space},  composed solely of $t$-AC states, have been studied in Ref.~\cite{Rudzinski2024orthonormalbasesof}.

There are equivalent characterizations of $(j,k,t)$ AC subspaces as specified in the following lemma.
 \begin{lemma}
\label{Lemma.Equiv.anti.sub}
     Let $V$ be a subspace of dimension $k$ of $\Hs^{(j)}$. The following statements are equivalent:
     \begin{enumerate}
         \item $V$ is a $(j,k,t)$ AC subspace.
         \item A (not necessarily orthogonal) basis $B_V=\{ \ket{\psi_m}\}_{m=1}^k$ of $V$ satisfies the following conditions:
         \begin{equation}
             \bra{\psi_{m_1}} T_{L M} \ket{\psi_{m_2}} = 0 \, ,  
         \end{equation}
         for any
         $m_1, \, m_2 = 1, \dots , k$, $L=1,\dots , t$ and $M= -L , \dots , L$.
         \item Any pair of states $\ket{\psi_1}, \ket{\psi_2} \in V$ satisfies the following conditions:
         \begin{equation}
             \bra{\psi_{1}} T_{L M} \ket{\psi_{2}} = 0 \, , 
         \end{equation}
         for any $L=1,\dots , t$ and $M= -L , \dots , L$.
     \end{enumerate}
 \end{lemma}
 
In particular, if $V$ is a $(j,k,1)$ AC subspace, we have that
\begin{equation}
\label{Eq.Cond.1AC.subspace}
\bra{\psi_{1}} J_a \ket{\psi_{2}} = 0 
\end{equation}
for any pair of states $\ket{\psi_{1}}, \ket{\psi_{2}} \in V$ and $a=x,y,z$ (including the case where $\ket{\psi_{1}}=\ket{\psi_{2}}$). One method of searching for AC subspaces, which is well suited to numerical approaches, is to define an objective function which takes as arguments \rema{$k$ linearly independent spin-$j$ spinors} and which is zero if and only if these spinors span a $(j,k,t)$ AC subspace for some $t$. To this end, we first introduce, for integers $L,M$, the functions
\begin{equation}
\label{Exp.plane}
        G_{L M}(B_V) = \sum_{m_1 \leq m_2}^k | \bra{\psi_{m_1}} T_{L M} \ket{\psi_{m_2}} |^2 \, ,
\end{equation}
where $B_V = \{ \ket{\psi_m} \}_{m=1}^k$ is now assumed to be an orthonormal basis of some subspace $V$ of dimension $k$. These functions can be equivalently rewritten as follows,
    \begin{equation}
    \label{Exp.DM}
        G_{L M}(\Pi_V) = \Tr \left( \Pi_V\, T_{L M} \,\Pi_V \,T_{L M}^{\dagger} \right)
    \end{equation}
highlighting the fact that they depend only on the projector onto $V$
\begin{equation}
    \label{DM}
       \Pi_V = \sum_{\ket{\psi_m}\in B_V} \ket{\psi_m} \bra{\psi_m}.
\end{equation}
The non-negative function $G_{LM}(\Pi_V)$ is zero if and only if all the terms in the sum~\eqref{Exp.plane} vanish simultaneously. Therefore, the  objective function 
\begin{equation}
\label{Obj.function}
G_t(\Pi_V) = \sum_{L=1}^t \sum_{M=-L}^{L} G_{L M}(\Pi_V) \, .
\end{equation}   
is zero if and only if $\bra{\psi_{m_1}} T_{LM} \ket{\psi_{m_2}} = 0$ for any $L=1,\dots , t$ and $M=-L, \dots , L$ and any $m_1 , m_2 = 1, \dots , k$, which corresponds to the second equivalent statement of a subspace $V$ being a $(j,k,t)$ AC subspace mentioned in Lemma~\ref{Lemma.Equiv.anti.sub}. In this work, most of the AC subspaces presented in Sections~\ref{Sec.OQR.small.spin}-\ref{Sec.Examples.OQR} were found by searching for zeros of the objective function $G_t(\Pi_V)$ by numerical optimization.
\subsection{Negativity of pure AC states}
\label{SubSec.Neg}
The negativity of a state $\rho$ of a bipartite system $AB$ with Hilbert space $\Hs^A \otimes \Hs^B$ is defined in terms of the eigenvalues $\La_k$ of the partial transpose of the state $\rho^{T_A}$ as~\cite{PhysRevA.58.883,PhysRevA.65.032314}
\begin{equation}
\label{Eq.Neg.Def}
\Nega_A (\rho) = 
 \sum_k \frac{|\La_k| -\La_k}{2} .
\end{equation}
The Schmidt decomposition of a pure $t$-AC state for the bipartition $A|B\equiv t | N-t$ (where we assume $t \leq N-t$ without loss of generality) is given by
\begin{equation}
\label{SDAC}
    \ket{\psi_{t\text{-}\mathrm{AC}}} = \frac{1}{\sqrt{t+1}} \sum_{\alpha=1}^{t+1} \ket{\phi^{A}_{\alpha}} \otimes \ket{\phi^{B}_{\alpha}} ,
\end{equation}
with $\{ \ket{\phi_{\alpha}^A}\}$ an orthonormal basis of $\Hs^A$, and $\{ \ket{\phi_{\alpha}^B}\}$ an orthonormal subset of $\Hs^B$. An immediate consequence of \eqref{SDAC} is that the negativity of any pure $t$-AC state $\rho_{\psi_{t\text{-}\mathrm{AC}}}=|\psi_{t\text{-}\mathrm{AC}}\rangle\langle \psi_{t\text{-}\mathrm{AC}}| $ is maximal and equal to~\cite{Den.Mar:22}
\begin{equation}
\label{Eq.Max.Neg.Pure}
\Nega_t(\rho_{\psi_{t\text{-}\mathrm{AC}}}) = \frac{t}{2} \, .    
\end{equation}
This maximization of negativity no longer applies to anticoherent mixed states, as shown by the example of the maximally mixed state $\rho_0$, which is $t$-AC to any order $t$ but for which $\Nega_t(\rho_0)=0$ because $\rho_0$ is fully separable. Consequently, for arbitrary mixed states, there is a priori no relationship between negativity and anticoherence. However, in Sec.~\ref{Sec.Neg.AC}, we will uncover a connection between $t$-AC subspaces and mixed states with maximal negativity over certain bipartitions.
\section{Optimal quantum rotosensors for infinitesimal rotations}
\label{Sec.OQR}
\subsection{\rema{Definition and characterization}}
We consider the problem of optimal quantum rotosensors (OQR)~\cite{Kolenderski_Demkowicz-Dobrzanski_2008,Chr.Her:17,Mar.Wei.Gir:20,ZGoldberg_2021} for infinitesimal rotations, i.e., for rotation angles $\eta \ll 1 $.  
\rema{An OQR, as defined in~\cite{Chr.Her:17}, is a state that minimizes the fidelity, averaged over the rotation axis, which, for $\eta \ll 1$, is equivalent to maximizing the average QFI (see Eqs.~\eqref{Eq.Fidelity.small}-\eqref{Def.QFI})
\begin{equation}
\label{Eq.QIF.ave}
\overline{I(\mathbf{n},\rho)} = 
\frac{1}{4\pi} \int 
I(\mathbf{n},\rho)
\diff \mathbf{n} .
\end{equation}
We will call this type of state \emph{OQR with respect to fidelity}, and denote it by $\OQRI$. Additionally, an alternative optimization is considered with respect to the best estimate of $\eta$, specifically the one provided by the QCRB in Eq.~\eqref{Eq.QCRB}. States that minimize the averaged QCRB
}
\begin{equation}
\label{QCRB.Eq}
 M\, \overline{\text{Var} [\hat{\eta}] } 
 \geqslant  \overline{I(\mathbf{n},\rho)^{-1}} \, ,
\end{equation}
\rema{i.e., those that minimize the average of the inverse of the QFI over all axes of rotation, will be called \emph{OQR with respect to the QCRB}, denoted by $\OQRInv$.} The r.h.s.\ of Eq.~\eqref{QCRB.Eq} is lower bounded by Jensen's inequality~\cite{durrett2019probability}
\begin{equation} 
\label{QCRB.Eq.Jensen}
\overline{I(\mathbf{n},\rho)^{-1}} 
\geqslant
\rema{ \frac{1}{\phantom{a} \overline{I(\mathbf{n},\rho)} \phantom{a}} }
 \, , 
\end{equation} 
which is tight when the function $I(\mathbf{n} , \rho)$ is independent of the variable of integration $\mathbf{n}$. Therefore, \rema{an $\OQRInv$ needs to saturate Jensen's inequality in addition to maximize $\overline{I(\mathbf{n},\rho)} $.} Therefore, we have two conditions that characterize the OQRs:
\begin{subequations}
\begin{flalign}
   & \bullet~~\rho~\textrm{maximizes}~\overline{ I(\mathbf{n} , \rho)}\phantom{--} \label{Cond.1} \\[5pt]
   & \bullet~~I(\mathbf{n} , \rho)~\textrm{is independent of}~\mathbf{n}\phantom{--} \label{Cond.2}
\end{flalign}
\end{subequations}
\rema{An $\OQRI$ satisfies condition~\eqref{Cond.1} while an $\OQRInv$ \rema{fulfills} both}. As we show below, these two conditions naturally lead to the notion of AC states and subspaces.

\rema{Before beginning the study of OQRs for pure and mixed states, let us comment on the differences between the two types of OQRs with respect to their associated measurement protocols. In both cases, an OQR is an optimal state for estimating an infinitesimal rotation angle when the rotation axis is unknown and uniformly distributed on the sphere. However, for an $\OQRI$, we consider a fidelity measurement protocol (see Ref.~\cite{Liang_2019} for a review) between the initial state and the rotated state to estimate the rotation angle. On the other hand, the ultimate estimate of an $\OQRInv$ is achievable in principle by optimizing over all Positive Operator Valued Measures (POVM)~\cite{RevModPhys.90.035005}. There are some explicit derivations of the optimal POVM in the literature, for the estimation of a general parameter~\cite{Bra.Cav:94,Hayashi5630,RevModPhys.90.035005}, and for the multiparametric estimation of rotations~\cite{ZGoldberg_2021,PhysRevApplied.20.024052}}.
\subsection{Pure spin-$j$ states}
For a pure state $\rho_{\psi} \equiv \ket{\psi} \bra{\psi}$, the Uhlmann-Jozsa fidelity~\eqref{Eq.Fidelity.mixed} reduces to~\eqref{Fidelity.pure.general}, and the QFI~\eqref{Def.QFI} is just the variance of the angular momentum component along $\mathbf{n}$~\cite{Bra.Cav:94}
\begin{equation}
\label{QFI.pure.states}
     I(\mathbf{n} , \rho_\psi )  = 4 \left( \langle J_{\mathbf{n}}^2 \rangle - \langle J_{\mathbf{n}} \rangle^2  \right)
\end{equation}
with $\langle A \rangle = \langle \psi | A |\psi \rangle $. The averaged QFI is then equal to
\begin{equation}
\label{Eq.AvI.pure}
\begin{aligned}    
 \overline{  I(\mathbf{n} , \rho_\psi) }
 ={}& 
 \frac{4}{3} \sum_{a=x,y,z} \Big( \langle J_a^2 \rangle - \langle J_a \rangle^2 \Big)\\
 ={}& \frac{4}{3} \Big(
j(j+1) - \langle \mathbf{J} \rangle^2
\Big)
\,,
\end{aligned}
\end{equation}
where we used the identity
\begin{equation}
\label{Eq.Id.Integral}
    \int_{S^2} n_a n_b \, \diff \mathbf{n} = \frac{4\pi}{3} \delta_{ab} 
\end{equation}
to evaluate the integral defining the average. The conditions \eqref{Cond.1}--\eqref{Cond.2} for a pure state $\rho_{\psi}$ are fulfilled whenever $\langle J_{\mathbf{n}} \rangle=0$ and $\langle J_{\mathbf{n}}^2 \rangle = j(j+1)/3$ for any $\mathbf{n}$, that is when $\ket{\psi}$ is a $2$-AC state~\cite{Gol.Jam:18,PhysRevLett.127.110501}. In this case, the optimal lower bound for the averaged QCRB reads
\begin{equation}
\label{QCRB.2ant}
\overline{\text{Var} [\hat{\eta}] }
\geqslant \frac{3}{4Mj(j+1)} ,
\end{equation}
that is inversely proportional to $j(j+1)$. This means that the higher the spin quantum number, the better the optimal estimate that can be obtained. 
\begin{table*}[t!]
    \centering
    \begin{tabular}{@{\hskip 0.2in} l @{\hskip 0.1in}| @{\hskip 0.5in}c@{\hskip 0.5in}c@{\hskip 0.5in} c}
 \rema{OQR}  & \rema{Conditions}    & Pure states $\ket{\psi}$ & Mixed states $\rho$
      \\[3pt]
      \hline \\[-10pt]     
      \rema{$\OQRI$} & \eqref{Cond.1} 
      & $\ket{\psi}$ is $1$-AC & $\IMA (\rho)$ is a $1$-AC subspace
       \\[0.12cm]
    \rema{$\OQRInv$} & \eqref{Cond.1} and \eqref{Cond.2}
    & $\ket{\psi}$ is $2$-AC 
    & \rema{$\IMA (\rho)$ is a $1$-AC subspace and $\rho$ is $2$-AC}
   \end{tabular}
    \caption{\rema{Conditions that characterize, for pure and mixed states, the two types of OQRs: with respect to the fidelity ($\OQRI$) and with respect to the QCRB ($\OQRInv$). For both pure and mixed $\OQRInv$, the averaged QCRB is equal to $3/4Mj(j+1)$.} 
    }
    \label{tab.OQR}
\end{table*}
\subsection{Mixed spin-$j$ states}
We consider now mixed spin-$j$ states $\rho \in \HSs (\Hs^{(j)})$. Let $\{ \la_m , \ket{\psi_m} \}_{m=1}^d$ denote the eigenvalues and the corresponding eigenvectors of $\rho$ with $d=2j+1$ the dimension of $\Hs^{(j)}$. Without loss of generality, we label by $m=1, \dots, k$ the eigenvectors with the non-zero eigenvalues, which span the image $\IMA (\rho)$ of $\rho$, with $\dim (\IMA (\rho))=k$. When the state $\rho$ is not full rank ($k< d$), the states $\ket{\psi_m}$ with $m=k+1,\dots, d$ span the kernel of $\rho$, $\ker (\rho)$. The subspaces $\IMA (\rho)$ and $\ker (\rho)$ are \rema{orthogonal} complementary subspaces in $\HSs (\Hs^{(j)})$. We denote by $P_{\rho}$ and $P_{\rho}^{\perp}$ the projectors onto $\IMA (\rho)$ and $\ker (\rho) = \IMA (\rho)^{\perp}$, respectively,
\begin{equation}
\begin{aligned}
    & P_{\rho} = \sum_{m=1}^k \ket{\psi_m}\bra{\psi_m} , \\ 
    & P_{\rho}^{\perp} = \mathds{1} - P_{\rho} = \sum_{m=k+1}^d \ket{\psi_m}\bra{\psi_m}.
\end{aligned}
\end{equation}
The QFI~\eqref{Def.QFI}, in terms of the Uhlmann-Jozsa fidelity~\eqref{Eq.Fidelity.mixed}, is now given, for an infinitesimal rotation about the axis $\mathbf{n}$, by~\cite{Bra.Cav:94,PhysRevX.1.021022,PhysRevLett.123.250502}
\begin{equation}
\label{QFI.mixed}
    I(\mathbf{n},\rho) = 2 \sum_{l,m=1}^{d} p_{lm} |\bra{\psi_l} J_{\mathbf{n}} \ket{\psi_m}|^2 
\end{equation}
with
\begin{equation}
\quad p_{lm} = \left\{ \begin{array}{cc}
         0 & \text{ if } \la_m=\la_l =0 \\[4pt]
        \frac{\left( \la_m - \la_l \right)^2}{\la_m + \la_l} &  \text{ otherwise }
    \end{array} \right. .   
\end{equation} 
We can now rewrite Eq.~\eqref{QFI.mixed} as follows
\begin{equation}
\label{QFI.mixed.2}
\begin{aligned}
    I (\mathbf{n}, \rho) = {}& 2 \sum_{l,m =1}^k p_{lm}  |\bra{\psi_l} J_{\mathbf{n}} \ket{\psi_m}|^2 + \, 4\,\Tr (\rho J_{\mathbf{n}} P_{\rho}^{\perp} J_{\mathbf{n}}) 
    \\
    = {}& 2 \sum_{l,m =1}^k p_{lm}  |\bra{\psi_l} J_{\mathbf{n}} \ket{\psi_m}|^2 + 4\, \Tr \left( \rho J_{\mathbf{n}}^2 \right)
    \\ & \quad\quad
    - 2 \sum_{l,m =1}^k \left( \la_l + \la_m \right)  |\bra{\psi_l} J_{\mathbf{n}} \ket{\psi_m}|^2 
    \\
    = {}& 
    4\, \Tr \left( \rho J_{\mathbf{n}}^2 \right)
    -8 \sum_{l,m=1}^k \frac{\la_l  \la_m}{\la_l + \la_m} |\bra{\psi_l} J_{\mathbf{n}} \ket{\psi_m}|^2
    \, .
\end{aligned}
\end{equation}
It is easy to see from the last equation that $I(\mathbf{n}, \rho)$ reduces to Eq.~\eqref{QFI.pure.states} when the state is pure (i.e., when it has only one non-zero eigenvalue equal to $1$). The QFI for a mixed state \eqref{QFI.mixed.2} averaged over all directions is then given by 
\rema{
\begin{equation}
\begin{aligned}    
\label{Eq.AvI.Mixed}
    & \overline{I(\mathbf{n} ,\rho)} \\
    & = 
    \frac{4}{3}
    \sum_{a} \bigg(\Tr \left( \rho J_{a}^2 \right) - \sum_{l,m=1}^k \frac{2 \la_l  \la_m}{\la_l + \la_m} |\bra{\psi_l} J_{a} \ket{\psi_m}|^2 \bigg)
\\    
    & = \frac{4}{3}\bigg( 
    j(j+1)
    -  \sum_{l,m=1}^k \frac{2 \la_l  \la_m}{\la_l + \la_m}  \sum_{a} |\bra{\psi_l} J_a \ket{\psi_m}|^2
    \bigg) ,
\end{aligned}
\end{equation}
where we have used Eq.~\eqref{Eq.Id.Integral} to evaluate the integrals.} Hence, the first condition~\eqref{Cond.1} defining a mixed OQR reduces to satisfying $\bra{\psi_l} J_a \ket{\psi_m}=0$ for $a=x,y,z$ and any pair of states in $\IMA (\rho)$, which is equivalent to $\IMA (\rho)$ being a $1$-AC subspace. This condition implies that $\rho$ is a $1$-AC state, but it is actually stronger than that. Once the first condition~\eqref{Cond.1} is met, the second condition~\eqref{Cond.2} reduces to $\Tr (\rho J_{\mathbf{n}}^2)$ being independent of $\mathbf{n}$, i.e., $\rho$ being a $2$-AC state. Remarkably, the mixed states satisfying these two conditions, which we now call \emph{mixed OQRs}, are as good estimators as the pure OQRs because they give the same lower bounds \rema{in both cases, $\OQRI$ and $\OQRInv$}~\eqref{QCRB.2ant}. We can therefore conclude that a lower purity spin state is not necessarily a worse estimator. 

We summarize the conditions we have found for a state to be \rema{ either $\OQRI$ or $\OQRInv$}, in Table~\ref{tab.OQR}, both for the pure and mixed cases. The condition~\eqref{Cond.2} can always be fulfilled for any value of the spin quantum number by sufficiently mixing a state with the maximally mixed state and thus decreasing its purity $P=\Tr (\rho^2)$ (see proof in Appendix~\ref{App.t.anticoherence}). On the other hand, the first condition is increasingly difficult to fulfill as the rank of the state $\rho$ increases. There is therefore a certain tension between the two conditions. Since $\IMA (\rho)$ is of dimension equal to the rank $k$ of $\rho$, condition~\eqref{Cond.1} will be harder to meet for highly mixed states. The rank $k$ of a state is in fact directly related to the minimal purity it can reach, since $\Tr (\rho^2) = \sum_{m=1}^k \lambda_m^2\geqslant 1/ k$. For a given spin quantum number $j$, the maximum dimension $k_{\max}$ of a $1$-AC subspace sets in turn a lower bound on the purity attainable by mixed OQRs. We have determined the value of $k_{\max}$ by numerically searching for zeros of the objective function \eqref{Obj.function} for $j\leqslant 17/2$ (see discussion in Sec.~\ref{Sec.Examples.OQR} and Fig.~\ref{fig.kmax.subspaces}, where we show the dependence of $k_{\max}$ on $j$). 
\rema{
\subsection{General spin states}
\label{SubSec.OQR.general}
Let us now consider the case of a general finite-dimensional quantum system on which an action of $\mathrm{SU}(2)$ is defined. A generic state of the system can always be written as
\begin{equation}
    \ket{\Psi} =  \sum_{j=1}^{j_{\max}} \sum_{\alpha_j=1}^{g_j}
    c_{(j,\alpha_j)} 
    \ket{\varphi_{(j,\alpha_j)}} \in \Hs =  \bigoplus_{j=1}^{j_{\max}}
    \bigoplus_{\alpha_j=1}^{g_j}
    \Hs^{(j,\alpha_j)} ,
\end{equation}
where the finite dimensionality of the system implies that there is a maximum spin value $j_{\max}$. The additional index $\alpha_j$ is introduced to distinguish multiple copies of $\Hs^{(j)}$, the latter appearing with multiplicity $g_j$ in the sum above. In this general case, the conditions for a state to be an OQR are exactly the same as in the scenario considered in the previous section. In particular, \rema{condition~\eqref{Cond.1}} implies that $\rho$ must be a convex combination of states belonging only to the $j_{\max}$-irreps, ensuring the maximization of the term in $j(j+1)$ in Eq.~\eqref{Eq.AvI.Mixed}. Consequently, the OQRs are states of the form
\begin{equation}
    \ket{\Psi} = \sum_{\alpha=1}^{g_{j_{\text{max}}}}  c_{\alpha} \ket{\varphi_{\alpha}} \;\in\; 
    \bigoplus_{\alpha}  \Hs^{(j_{\max},\alpha)}
    \, \xhookrightarrow{} \, 
    \Hs .
\end{equation}
When $g_{j_{\max}}=1$, the OQRs reduce to the same states as those found in a single spin-$j_{\max}$ system. This case arises for both symmetric (bosonic) and antisymmetric (fermionic) qudit systems~\cite{Chr.etal:21}.
For systems with $g_{j_{\max}}>1$, on the other hand, OQRs can be systematically constructed as convex combinations of OQRs defined within each copy of $\Hs^{(j_{\max})}$ separately. Consequently, all the examples of OQRs given for single-spin systems can be generalized to any finite-dimensional system.  
}
\subsection{Comparison with the estimation of a rotation around a fixed axis}
If the axis of rotation is fixed and known, the optimal state must maximize $I(\mathbf{n},\rho)$, where we can take $\mathbf{n}=\mathbf{e}_z$ without loss of generality. We obtain from Eq.~\eqref{QFI.mixed.2} that the maximum of the QFI for all mixed states of rank $k$ with a fixed spectrum $\la_1 \geqslant \la_2 \geqslant \dots \geqslant \la_{k}$ is
\begin{equation}
\max I(\mathbf{e}_z,\rho) = 4\, \sum_{l=1}^k \la_l  \left( j- \left\lfloor\tfrac{l-1}{2} \right\rfloor \right)^2 ,
\end{equation}
which is achieved when the eigenstates of $\rho$ are given by
\begin{equation}
    \ket{\psi_l} = N_l \Big( \left| j,j- \left\lfloor \tfrac{l-1}{2} \right\rfloor \right\rangle +(-1)^{l-1} \left| j,\left\lfloor \tfrac{l-1}{2} \right\rfloor - j \right\rangle \Big) , 
\end{equation}
with normalization factors
\begin{equation}
N_l = \left\{ \begin{array}{cc}
     1/2 & \text{ if } j \in \mathds{Z} \text{ and } l=2j+1
      \\      
      1/\sqrt{2} &  \text{ otherwise }
\end{array}
\right.  .
\end{equation}
The latter are also eigenstates of $J_z^2$ with eigenvalues $h_l = (j- \left\lfloor (l-1)/2 \right\rfloor)^2 $.
The same optimal mixed state can be also deduced from the results of Ref.~\cite{PhysRevLett.123.250502}. We observe that if the state has a rank less than or equal to 2, $I(\mathbf{e}_z,\rho)=4j^2$ (see also~\cite{PhysRevLett.123.250502,PhysRevX.1.021022}). However, for mixed states of rank $k>2$, it is inevitable that the QFI decreases and, consequently, that the resulting state is less sensitive than the pure optimal state. On the contrary, and as we will show in the following, there are mixed OQRs of higher ranks with the same QCRB as pure OQRs for a sufficiently large spin.
\section{Mixed OQRs for small spin}
\label{Sec.OQR.small.spin}
\rema{We present here several examples of mixed OQRs.} As in the case of pure states~\cite{Chr.Her:17,Mar.Wei.Gir:20,ZGoldberg_2021}, the existence of OQRs is linked to that of AC states.  
\rema{
A general approach to identify mixed OQRs is to find first $1$-AC subspaces $V$ which can be used to construct mixed states satisfying the condition~\eqref{Cond.1}, followed by the search for an appropriate mixture of states within $V$ that will satisfy the condition~\eqref{Cond.2}. Another, more direct approach is to find $2$-AC subspaces from which a mixture of basis states will automatically define a mixed OQR. In both cases, it is essential to search for and identify AC subspaces.}
\subsection{Spin 1}
The only $1$-AC spin-$1$ states are given by $\Rr^{(1)}\ket{1,0}$~\cite{Bag.Bas.Mar:14}, where $\Rr^{(1)} = \Rr^{(1)} (\alpha,\beta, \gamma)$ is the rotation matrix in the spin-1 irrep, parametrized by the Euler angles. First, let us search for a $(1,2,1)$ AC subspace spanned by an orthonormal basis $\{ \ket{\psi_1}, \ket{\psi_2} \}$. The two states $\ket{\psi_a}$ must be obtained from the state $\ket{1,0}$ by rotations, since they must be $1$-AC. Moreover, since the order of anticoherence is independent of the orientation of the subspace, we can reorient the basis so that
\begin{equation}
\begin{aligned}    
    \ket{\psi_1}= {}& \ket{1,0} = (0,1,0) \, , 
    \\
    \ket{\psi_2} = {}& \Rr^{(1)}(\alpha, \beta, \gamma) \ket{1,0} = \left( -\tfrac{e^{-i \alpha } \sin \beta }{\sqrt{2}},\cos \beta ,\tfrac{e^{i \alpha } \sin \beta }{\sqrt{2}} \right)
    \end{aligned}
\end{equation}
where the \rema{spinor components refer to} the $\ket{j,m}$ basis, with $m=j,\dots, -j$. By direct calculation, we obtain $\langle \psi_2 | \psi_1 \rangle = \cos \beta$ and $\bra{\psi_2} J_{+} \ket{\psi_1} = e^{-i \alpha} \sin \beta$, which cannot be made simultaneously zero. 
Consequently, there are no $(1,k,1)$ AC subspaces with $k>1$. On the other hand, the only $2$-AC spin-$1$ state is the maximally mixed state $\rho_0 = \mathds{1}/3$. Therefore, \rema{there are only $\OQRI$ for pure states, but no $\OQRInv$ for either pure or mixed states of spin $j=1$}. 
\subsection{Spin 3/2}
As for spin 1, there is only one $1$-AC state, up to rotation, given by the GHZ state $\ket{\psi_{\text{GHZ}}}= (1,0,0,1 )/\sqrt{2}$~\cite{Bag.Bas.Mar:14}.  We search for a $(3/2,2,1)$ AC subspace spanned by a reoriented orthonormal basis such that
\begin{equation}
\begin{aligned}    
    \ket{\psi_1}=   \ket{\psi_{\text{GHZ}}} \, , \quad  
    \ket{\psi_2} =  \Rr^{(3/2)} (\alpha , \beta , \gamma)\ket{\psi_1}  \, .
    \end{aligned}
\end{equation}
The orthogonality condition $\langle \psi_2 | \psi_1 \rangle = 0$ has three solutions: (i) $\beta=0$ and $\alpha + \gamma = (2n_1+1)\pi/3$, (ii) $\beta=\pi$ and $\alpha - \gamma = 2n_1\pi/3$, and (iii) $\al = \pi (2n_1+2n_2+1)/6$ and $\gamma = \pi (2n_1 - 2n_2+1)/6$ for some integer numbers $n_1$ and $n_2$. The three solutions have the same general expression $\ket{\psi_2} = (a,b,c,-a)$ with $a,b,c\in\mathbb{C}$ and $a \neq 0$. By direct calculation, we get that $\bra{\psi_2} J_z \ket{\psi_1} \neq 0$ and we conclude that there are no $(3/2,2,1)$ AC subspaces, and thus no mixed OQRs, \rema{either $\OQRI$ or $\OQRInv$,} in this case.

In order to study the effect on the QCRB of a deviation from the condition \eqref{Cond.1} while satisfying the condition \eqref{Cond.2}, let us consider a family of $2$-AC mixed spin-$3/2$ states in the purity range $1/4 \leqslant \Tr (\rho^2) \leqslant 1/2$, as derived in Appendix~\ref{2Ant.3h2spin}. In particular, we show there that the $2$-AC mixed state with the largest purity $P=1/2$ is the rank-two state $\rho_{\text{$2$-AC}}= (\rho_{\psi_1}+\rho_{\psi_2})/2$ with
\begin{equation}
\label{St.s3h2}
    \ket{\psi_1} = \frac{1}{\sqrt{2}} (1,0,1,0) \, ,
    \quad
    \ket{\psi_2} = \frac{1}{\sqrt{2}} (0,-1,0,1).
\end{equation}
As we have already shown above,  $\IMA(\rho_{\text{$2$-AC}})$ does not span a $1$-AC subspace.

Despite the absence of \rema{$\OQRInv$}, we can compare the QCRBs~\eqref{QCRB.Eq} achieved by mixed and pure states, respectively. We have that
\begin{equation}
   \underbrace{\vphantom{\frac{3}{j(j+1)}}\overline{I(\mathbf{n},\rho_{\text{$2$-AC}})^{-1}}}_{=~0.25} \,\geqslant\,
   \underbrace{\vphantom{\frac{3}{j(j+1)}}\overline{I(\mathbf{n},\rho_{\text{GHZ}})^{-1}}}_{\approx~0.225}
   \, \geqslant\,
   \underbrace{\frac{3}{j(j+1)}}_{=~0.2},
\end{equation}
which means that the pure GHZ state is only slightly better than the $2$-AC mixed state with the highest purity $P=1/2$.
\subsection{Spin 2}
\label{sec:j2}
\begin{figure}[t!]
    \includegraphics[width=0.435\textwidth]{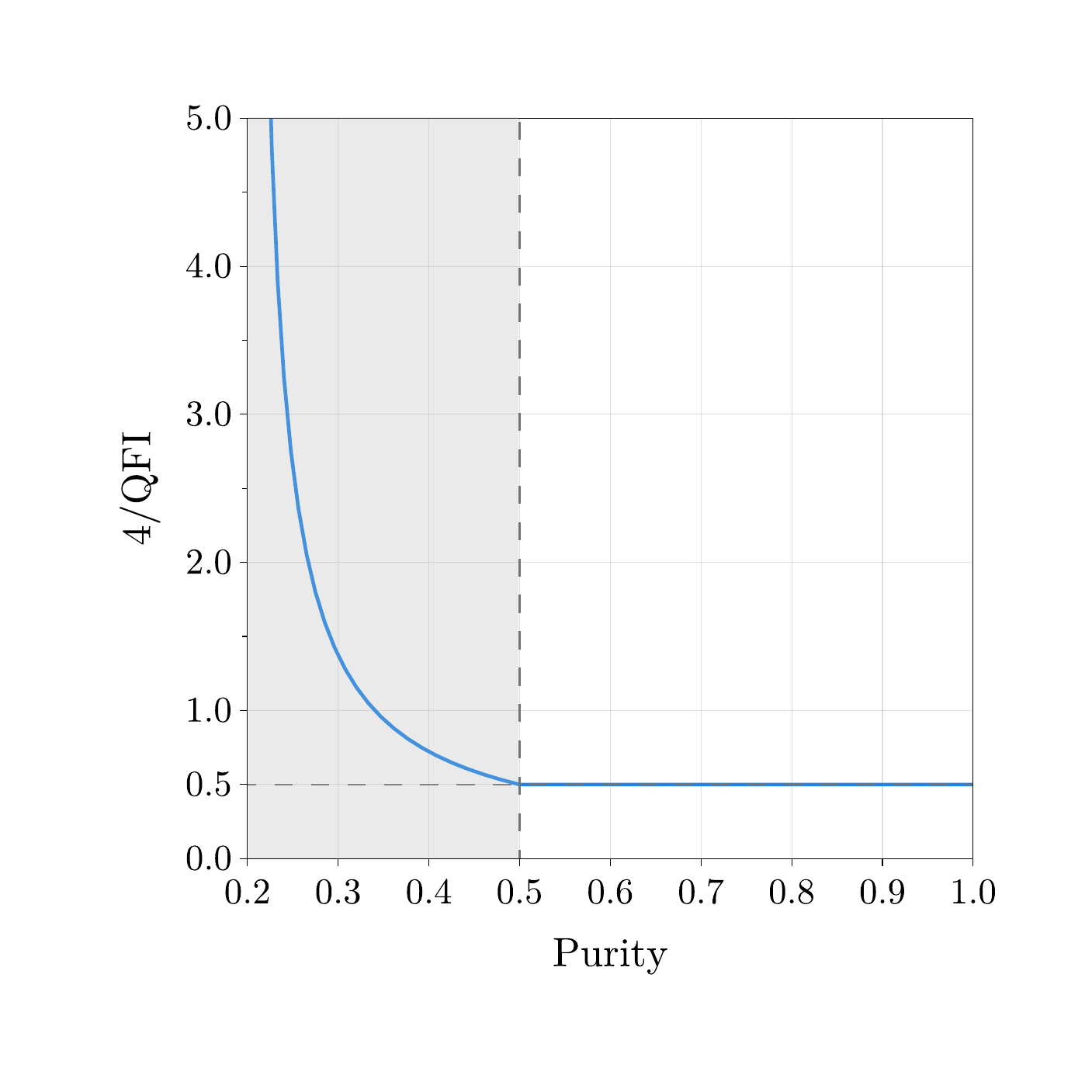}
    \caption{\label{Fig1}
     QCRB as a function of the purity $P(\rho(\xi))$ for the family of mixed spin-$2$ states~\eqref{Fam.spin2}. For the whole family, $I(\mathbf{n},\rho(\xi))^{-1}$ is independent of $\mathbf{n}$. States with a purity $P\geqslant 1/2$ are $\OQRInv$. For $P < 1/2$ (grey shaded region), the QCRB increases as the purity decreases, until it reaches infinity for the minimum purity $0.2$ (that of the maximally mixed state).
    }
\end{figure}

This is the smallest spin quantum number for which there is a two-dimensional $1$-AC subspace. Such a subspace is spanned by the orthogonal pure states
\begin{equation}
\begin{aligned}
\label{j2k2t1.1stplane}
    & \ket{\psi_1} = \frac{1}{2} \left( 1 , 0, \sqrt{2}\, i , 0 ,1\right),\\
    & \ket{\psi_2} = \frac{1}{2} \left( 1 , 0, -\sqrt{2}\, i , 0 ,1\right) .
\end{aligned}
\end{equation}
Since both states $\ket{\psi_k}$ are $2$-AC~\cite{Bag.Dam.Gir.Mar:15,Bag.Mar:17}, so is any mixture $\rho = \la_1 \ket{\psi_1}\bra{\psi_1}+ (1-\la_1) \ket{\psi_2}\bra{\psi_2}$, which therefore defines a mixed OQR. By varying the eigenvalues of $\rho$ through $\la_1$, we can obtain mixed OQRs with purity $\Tr (\rho^2) \in [1/2 , 1]$. 
We prove in Appendix~\ref{App.uniqueness.spin2} that the $(2,2,1)$ AC subspace spanned by the states~\eqref{j2k2t1.1stplane} is unique up to a rotation. As a result, spin-$2$ mixed OQRs are all of the above form. For mixed states of higher rank and smaller purity, the QCRB increases as the purity decreases. \rema{This tendency is exemplified} by considering the uniparametric family of mixed states
\begin{equation}
\label{Fam.spin2}
    \rho(\xi) = \left\{
\begin{array}{cr}
  \xi \rho_{\psi_1} + (1-\xi)\rho_{\psi_2} ,   
  & \xi \in \left[ \frac{1}{2} , 1 \right]
  \\[0.2cm]
 \left(\frac{5\xi-1}{3} \right)(\rho_{\psi_1} + \rho_{\psi_2}) + \left(\frac{1-2\xi}{3}\right)\mathds{1} , & \xi \in \left[ \frac{1}{5} , \frac{1}{2} \right] 
\end{array}
    \right. ,
\end{equation}
where the $\ket{\psi_k}$ are given in Eq.~\eqref{j2k2t1.1stplane}. For $\xi \in \left[ \frac{1}{2} , 1 \right]$, $\rho(\xi)$ is an OQR, but not for 
$\xi \in \left[ \frac{1}{5} , \frac{1}{2} \right]$. The purity of $\rho(\xi)$ increases monotonically with $\xi$ as
\begin{equation}
    P(\rho(\xi)) = \left\{
\begin{array}{cc}
1+  2 \xi(\xi-1)   & \xi \in \left[ \frac{1}{2} , 1 \right] 
  \\[0.2cm]
\frac{1}{3} [ 2 \xi (5 \xi-2)+1 ]  & \xi \in \left[ \frac{1}{5} , \frac{1}{2} \right] \end{array}
    \right.  .
\end{equation}
We also find that the QFI for $\rho(\xi)$ is independent of $\mathbf{n}$ and its inverse is equal to
\begin{equation}
    I(\mathbf{n},\rho(\xi))^{-1} = \left\{
\begin{array}{cc}
  \frac{1}{8}   & \xi \in \left[ \frac{1}{2} , 1 \right] 
  \\[0.2cm]
\frac{3 (\xi+1)}{16 (1-5 \xi)^2}    & \xi \in \left[ \frac{1}{5} , \frac{1}{2} \right] 
\end{array}
    \right. \, . 
\end{equation}
In Fig.~\ref{Fig1}, we plot $I(\mathbf{n},\rho(\xi))^{-1}$ as a function of $P(\rho(\xi))$. We see that, for $P <\frac{1}{2}$, the inverse of the QFI starts to increase until infinity is reached for the maximally mixed state ($\xi=1/5$). Therefore, the QCRB increases (and thus the metrological power of the state decreases) as the purity decreases and is below $1/2$, as claimed above.
\section{Examples of AC subspaces and mixed OQRs}
\label{Sec.Examples.OQR}
\rema{
In this section, we summarize several of our results on the existence of 1- and $2$-AC subspaces found numerically based on the objective function~\eqref{Obj.function} and on the possibility of obtaining a mixed OQR from them. We start with specific values of $(j,k,t)$, and then introduce families of AC subspaces of an arbitrary spin quantum number.}
\subsection{$(j,k,1)$ subspaces}

\rema{We found several non-rotationally-equivalent $1$-AC subspaces of dimension $2$ for $j=5/2$ (see Appendix~\ref{Appendix1ACsub}). A statistical mixture of two basis states for such subspaces is $1$-AC by construction and a candidate for a mixed $2$-AC state. However, we find that none of these mixed states is actually $2$-AC, satisfying the condition~\eqref{Cond.2}, even though they come very close for an equally weighted mixture. For example, we get a maximum of $\mathcal{A}_2= 63/64 \approx 0.9843$ based on the subspace $V_1$ given in Eq.~\eqref{Ex1.N5k2t1} and a maximum of $\mathcal{A}_2 =     
\frac{3}{32} \left(7 \sqrt{7}-8\right) \approx 0.9863$ based on the subspace  $V_2$ given in Eq.~\eqref{Ex2.N5k2t1}. The corresponding states are therefore not mixed OQRs. Although there are mixed $2$-AC states (see Appendix~\ref{App.t.anticoherence}), we observe numerically that none of them has $\IMA(\rho)$ forming a $1$-AC subspace. \rema{Therefore, there are $\OQRI$'s for spin-5/2, but we could not identify a mixed $\OQRInv$.}}

\rema{The first $1$-AC subspace $V$ of dimension $3$ appears for $j=3$ (see Eq.~\eqref{j3k3t1.Ex1}). No pure state $|\psi\rangle\in V$ is $2$-AC. On the other hand, a statistical mixture of the basis states of $V$ with appropriate weights can give a $\rho$ which is $2$-AC and hence a mixed \rema{$\OQRInv$}. It should be noted that this is the first example of a mixed \rema{$\OQRInv$} with no pure OQR in its image $\IMA(\rho)$. We even found a whole family of such OQRs, given by Eq.~\eqref{Eq.331.mixedOQR}. Finally, we present a $(9/2,4,1)$ AC subspace in Eq.~\eqref{1AC4planej9o2}.}
\subsection{$(j,k,2)$ subspaces}
\rema{The $2$-AC subspaces $V$ of dimension $k>1$ are particularly interesting because any convex combination of any orthonormal basis states of $V$ results in a mixed $\rema{\OQRInv}$. We found the first instance of a $2$-AC subspace of dimension $2$ for $j=7/2$ (see Eq.~\eqref{Ex.j7h222.subspace}). For spins $j=4$ and $j=9/2$, no such AC subspaces were found numerically. The next example we found is for $j=5$, again a 2-dimensional subspace (see Eq.~\eqref{Ex.522.subspace}). Based on our numerical results presented in Fig.~\ref{fig.kmax.subspaces}, it is plausible that $(j,2,2)$ subspaces exist for all $j>5$. Additionally, a $2$-AC subspace of dimension $3$ was first identified for spin $7$ (see Eq.~\eqref{Ex.j732.subspace}).}
\subsection{AC subspaces of arbitrary dimension}
\begin{figure}[t]
    \centering
    \includegraphics[width=0.475\textwidth]{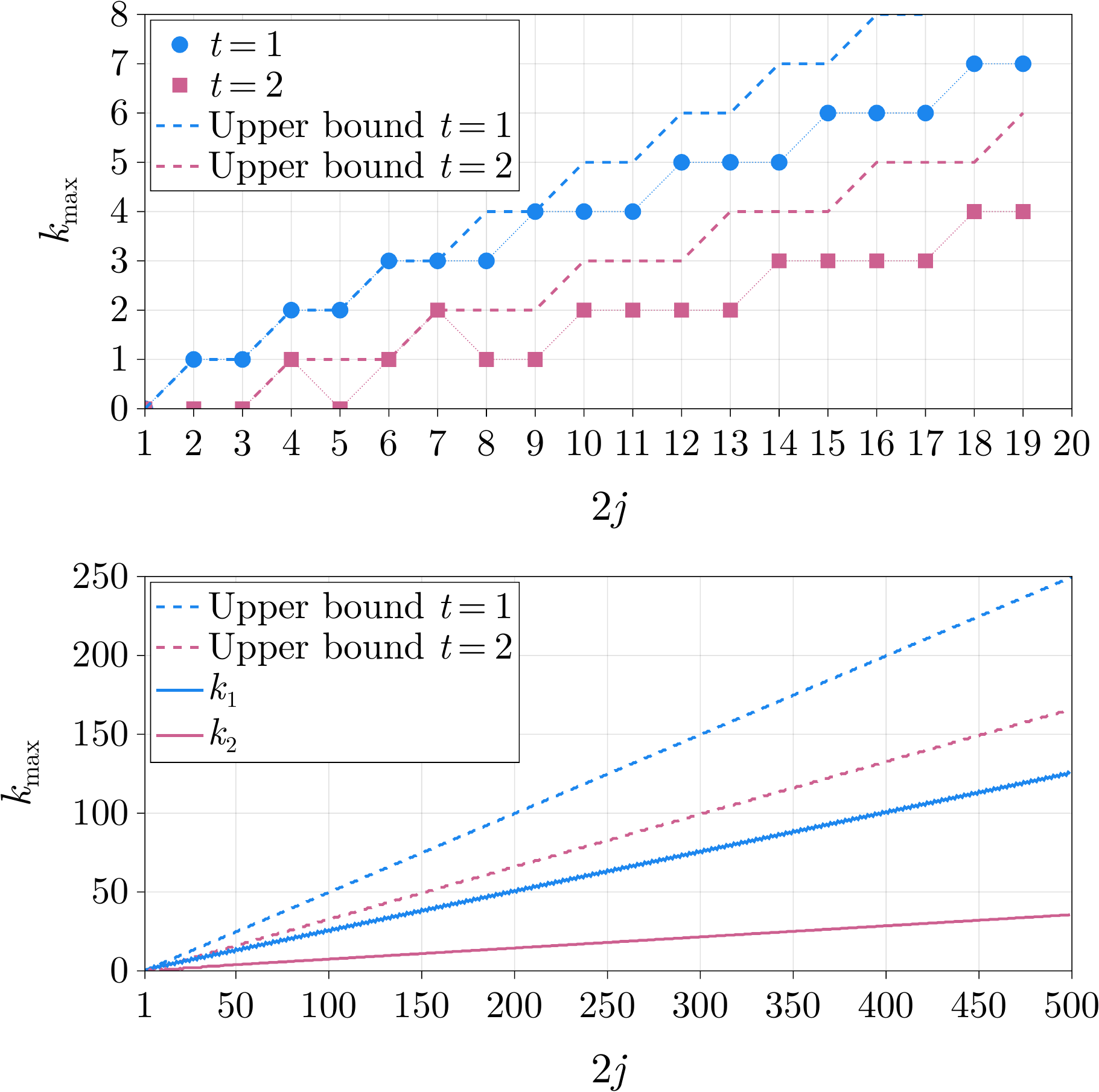}
    \caption{Top panel: Maximum dimension $k_\mathrm{max}$ of $(j,k,t)$ AC subspaces (determined numerically) as a function of the spin $j$ for anticoherence orders $t=1$ (blue dots) and 2 (pink squares). The dashed lines indicate the upper bound~\eqref{Eq.Upper.bound.kmax} for $t=1,2$. Bottom panel: Upper bound \eqref{Eq.Upper.bound.kmax} on $k_\mathrm{max}$ for $t=1$ and $t=2$ as a function of $j$ over a wide range of values for $j$. The dimensions of the $(j,k,t)$ AC subspaces defined in Eqs.~\eqref{Eq.1AC.sub1}--\eqref{Eq.1AC.sub2} and~\eqref{Eq.2AC.general} are also shown as solid lines.}
\label{fig.kmax.subspaces}
\end{figure}

Previously, we used the 1- and $2$-AC subspaces to create families of mixed OQRs. 
Here, we take a different approach and show that one can construct AC subspaces of order $t=1$ and $2$ of any dimension $k$ with sufficiently large spin using techniques similar to those outlined in Ref.~\cite{Bag.Dam.Gir.Mar:15}. 
For a spin-$j$ system with $j$ a half-integer number, the subspace spanned by the vectors
\begin{equation}
\label{Eq.1AC.sub1}
        \ket{\psi_a}= \frac{1}{\sqrt{2}}\Big( 
        \mathbf{0}_{2a} , 1 , \mathbf{0}_{2j-4a-1} , 1 , \mathbf{0}_{2a}
        \Big) 
\end{equation}
 with $a=0 , 1 , \dots, \lfloor \frac{j-1}{2} \rfloor$ is a $(j,k,1)$ AC subspace. Here, $\mathbf{0}_{n}$ is a string of $n$ zero components, with the convention that $\mathbf{0}_{0}$ means we do not add any extra zeroes. In the case where $j$ is an integer and $2j-3-4 \lfloor \frac{j-1}{2} \rfloor >0$, we can increase the dimension of the $1$-AC subspace by one by including the following additional state
\begin{equation}
\label{Eq.1AC.sub2}
\left( 
\mathbf{0}_j , 1 , \mathbf{0}_j
\right) \, .
\end{equation}
The dimension of these generic $1$-AC subspaces, denoted by $k_1$, is given by
\begin{equation}
k_{1}= \left\{ 
\begin{array}{cc}
     \left\lfloor \frac{j-1}{2} \right\rfloor +1 &  \text{for half-integer } j 
     \\[4pt]
     \left\lfloor \frac{j-1}{2} \right\rfloor +2 &  \text{for integer } j
\end{array}
\right.  \, ,
\end{equation}
and grows linearly as $j/2$ (see, e.g., Fig.~\ref{fig.kmax.subspaces}). As an illustration, we write the vectors spanning the $(j,k_1,1)$ AC subspaces for $j=2,7/2,4$ in Table~\ref{tab:1AC.subspaces}. The zero components in the states ensure that $\bra{\psi_a} T_{1 \pm 1} \ket{\psi_{a'}} \propto \bra{\psi_a} J_{\pm} \ket{\psi_{a'}} = 0 $ for any pair of states, where we used Eq.~\eqref{Eq.multipole.and.angular.momentum}. On the other hand, $\bra{\psi_a} T_{1 0} \ket{\psi_{a'}} \propto \bra{\psi_a} J_{z} \ket{\psi_{a'}} = 0 $ because $\ket{\psi_a} = ( \ket{j,m_a} + \ket{j,-m_a})/\sqrt{2} $ with $m_a \neq m_{a'}$ for $a \neq a'$.
\begin{table}[t!]
    \centering
    \begin{tabular}{c|l}
        Spin $j$~ & ~Basis of a $(j,k,1)$ AC subspace  \\
        \hline\\[-8pt]
         2 & $\left\{ \begin{array}{l}
              \frac{1}{\sqrt{2}} (1 , \mathbf{0}_3 ,1) \\
             ( \mathbf{0}_2 ,1 , \mathbf{0}_2 )
        \end{array} \right.$ 
        \\[16pt]
        7/2 & $\left\{ \begin{array}{l}
             \frac{1}{\sqrt{2}}(1 , \mathbf{0}_6 ,1) \\
             \frac{1}{\sqrt{2}}( \mathbf{0}_2 ,1 , \mathbf{0}_2 ,1 , \mathbf{0}_2 )
        \end{array} \right.$
        \\[16pt]
        4 & $\left\{ \begin{array}{l}
             \frac{1}{\sqrt{2}}(1 , \mathbf{0}_7 ,1) \\
             \frac{1}{\sqrt{2}}( \mathbf{0}_2 ,1 , \mathbf{0}_3 ,1 , \mathbf{0}_2 ) \\
             ( \mathbf{0}_4 ,1 , \mathbf{0}_4 )
        \end{array} \right.$
    \end{tabular}
    \caption{Table of some $1$-AC subspaces defined in Eqs.~\eqref{Eq.1AC.sub1}--\eqref{Eq.1AC.sub2} for spin $j=2,7/2,4$. Here, $\mathbf{0}_n$ denotes a string of $n$ zeros.}
    \label{tab:1AC.subspaces}
\end{table}

In the same way, it is possible to construct a generic $(j,k_2,2)$ AC subspace for $j\geq 5$ of dimension 
\begin{equation}
\label{Eq.Dimension.Gen.2AC.subspace}
k_2= \min \left( \left\lfloor \frac{j-\kappa-3}{3} \right\rfloor    , \frac{\kappa-2}{3} \right) +1  \, , 
\end{equation}
with 
\begin{equation}\label{kappa}
    \kappa = \left\lceil j - \sqrt{\frac{j(j+1)}{3}} \, \right\rceil \, ,
\end{equation}
spanned by the states
\begin{equation}
\label{Eq.2AC.general}
    \ket{\psi_a} = \left( \mathbf{0}_{3a} , \alpha_a , \mathbf{0}_{\kappa} , \beta_a , \mathbf{0}_{2j-3-2\kappa-6a} , \beta_a , \mathbf{0}_{\kappa} , \alpha_a , \mathbf{0}_{3a} \right) ,
\end{equation}
with $a=0, \dots, k_2-1$. Here again, the zero components of the states guarantee that $\bra{\psi_a} T_{\si \mu} \ket{\psi_{a'}} = 0$ for every $\si=1,2$, any $\mu \neq 0$, and any pair of states $\ket{\psi_a},\ket{\psi_{a'}}$. This is because the respective $T_{\si \mu}$ with $\mu\neq 0$ contains one or two ladder operators~\eqref{Eq.multipole.and.angular.momentum}. In addition, the zero state components guarantee that $\bra{\psi_a} T_{\si 0} \ket{\psi_{a'}}=0$ for different states. On the other hand, $\bra{\psi_a} T_{1 0} \ket{\psi_{a}}=0$, with $T_{10} \propto J_z$, is fulfilled by the property that the coefficients $\alpha_a$ and $\beta_{a}$ are associated with eigenstates $\ket{j, \pm m}$ whose magnetic quantum number is opposite. Finally, it is possible to satisfy the equality $\bra{\psi_a} T_{20} \ket{\psi_a}=0$, where $T_{20}\propto j(j+1)\mathds{1} - 3 J_z^2$ by choosing the parameters $\alpha_a$ and $\beta_a$, which we can considered as positive real numbers, such that $2\alpha_a^2 + 2\beta_a^2 =1$ (normalization of the state) and
\begin{equation*}
    j(j+1) - 6 \left[  (j-3a)^2\alpha_a^2 + (j-3a-1 -\kappa)^2\beta_a^2  \right] = 0.
\end{equation*}
Indeed, the above equation always has a solution for the value of $\kappa$ defined in Eq.~\eqref{kappa}. The dimension of the $(j,k_2,2)$ AC subspace~\eqref{Eq.2AC.general} grows asymptotically as $k_2 \approx (\sqrt{3}-1)j/(3\sqrt{3}) $. Some examples of $(j,k,2)$ AC subspaces constructed using this approach are given in Table~\ref{tab:2AC.subspaces}.

In Fig.~\ref{fig.kmax.subspaces} we show the dimension of the $1$- and $2$-AC subspaces given by our construction as a function of the spin quantum number $j$.  We note that both grow \rema{linearly} in the asymptotic limit $j\gg 1$. We also plot the largest dimension $k_{\mathrm{max}}$ of a $(j,k,t)$ AC subspace found numerically for $t=1,2$ and $j \leqslant 17/2 $. The value $k_{\mathrm{max}}$ for $1$-AC subspaces puts a lower bound on the purity of optimal OQRs, $\Tr (\rho^2)\geqslant 1/k_{\max}$. \rema{Thus, the larger $k_{\mathrm{max}}$ is, the more likely it is to find mixed OQRs of lower purity.} Our numerical results show that the Corollary 2 in Ref.~\cite{Per.Pau:17}, stating that the largest $1$-AC subspace of spin $j$ states has dimension $\lfloor j+1 \rfloor$, is not tight.

\begin{table}[t!]
    \centering
    \begin{tabular}{c|l}
        Spin $j$~ & ~Basis of a $(j,k,2)$ AC subspace  \\
        \hline\\[-8pt]
         5 & $
             \left( \frac{1}{\sqrt{7}} , \mathbf{0}_2 , \sqrt{\frac{5}{14}} , \mathbf{0}_3 , \sqrt{\frac{5}{14}}  , \mathbf{0}_2 , \frac{1}{\sqrt{7}}  \right)
        $ 
        \\[12pt]
        11 & $\left\{ \begin{array}{l}
             \frac{1}{8 \sqrt{3}}\left( \sqrt{19} , \mathbf{0}_5 , \sqrt{77} , \mathbf{0}_9 , \sqrt{77}  , \mathbf{0}_5 , \sqrt{19} \right) \\
             \frac{1}{\sqrt{6}} \left( \mathbf{0}_3 , \sqrt{2} , \mathbf{0}_5 , 1 , \mathbf{0}_3 , 1 , \mathbf{0}_5 , \sqrt{2} , \mathbf{0}_3  \right)
        \end{array} \right.$
        \\[16pt]
        35/2 & $\left\{ \begin{array}{l}
             \frac{1}{6\sqrt{39}} \left( \sqrt{107} , \mathbf{0}_8 , \sqrt{595} , \mathbf{0}_{16} , \sqrt{595}  , \mathbf{0}_8 , \sqrt{107} \right) \\
             \frac{1}{6\sqrt{30}} \left( \mathbf{0}_3 ,\sqrt{233} , \mathbf{0}_8 , \sqrt{307} , \mathbf{0}_{10} , \sqrt{307} , \mathbf{0}_8 ,\sqrt{233}  ,  \mathbf{0}_3 \right) \\
             \frac{1}{6\sqrt{21}} \left( \mathbf{0}_6 , \sqrt{305} , \mathbf{0}_8 , \sqrt{73} , \mathbf{0}_4 , \sqrt{73} , \mathbf{0}_8 , \sqrt{305} , \mathbf{0}_6 \right)
        \end{array} \right.$
    \end{tabular}
    \caption{Table of some $(j,k,2)$ AC subspaces defined in Eq.~\eqref{Eq.2AC.general} for spin $j=5,11,35/2$. Here, $\mathbf{0}_n$ denotes a string of $n$ zeros.}
    \label{tab:2AC.subspaces}
\end{table}

\section{Negativity and anticoherent subspaces}
\label{Sec.Neg.AC}
The quantum enhancement in metrological power is generally attributed to entanglement \rema{of a multipartite system, or to other quantum correlations for single-particle systems}. \rema{In this section, we study}
the relationship between the negativity $\Nega_t(\rho)$ of a \rema{spin-$j$} mixed \rema{OQR} $\rho$, \rema{seen as a symmetric state of $N=2j$ qubits}, and its AC properties.  
Despite the classical character of a mixture of states, we will show that a mixed state such that $\IMA(\rho)$ is a $(j,k,t)$ AC subspace maximizes all the negativities $\Nega_{t'}(\rho)$ for $t'\leqslant t$. In particular, the negativity of these mixed states with respect to the bipartitions $t'|N-t'$ with $t'\leqslant t$ is equal to that of the most entangled pure states~\eqref{Eq.Max.Neg.Pure}. Because of the relevance of this result, we state it as a theorem, the proof of which appears in Appendix~\ref{App.Neg.Sub}:
\begin{theorem}
\label{Theorem.1}
    Let $\rho \in \HSs (\Hs^{(j)})$ be a mixed state such that $\IMA (\rho)$ is a $(j,k,t)$ AC subspace. Then $\rho$, seen as a $N$-qubit symmetric state, maximizes the negativity with respect to the bipartition $t|N-t$ with $\Nega_t(\rho)= t/2$.
\end{theorem}
\begin{figure}[t!]
\includegraphics[width=0.475\textwidth]{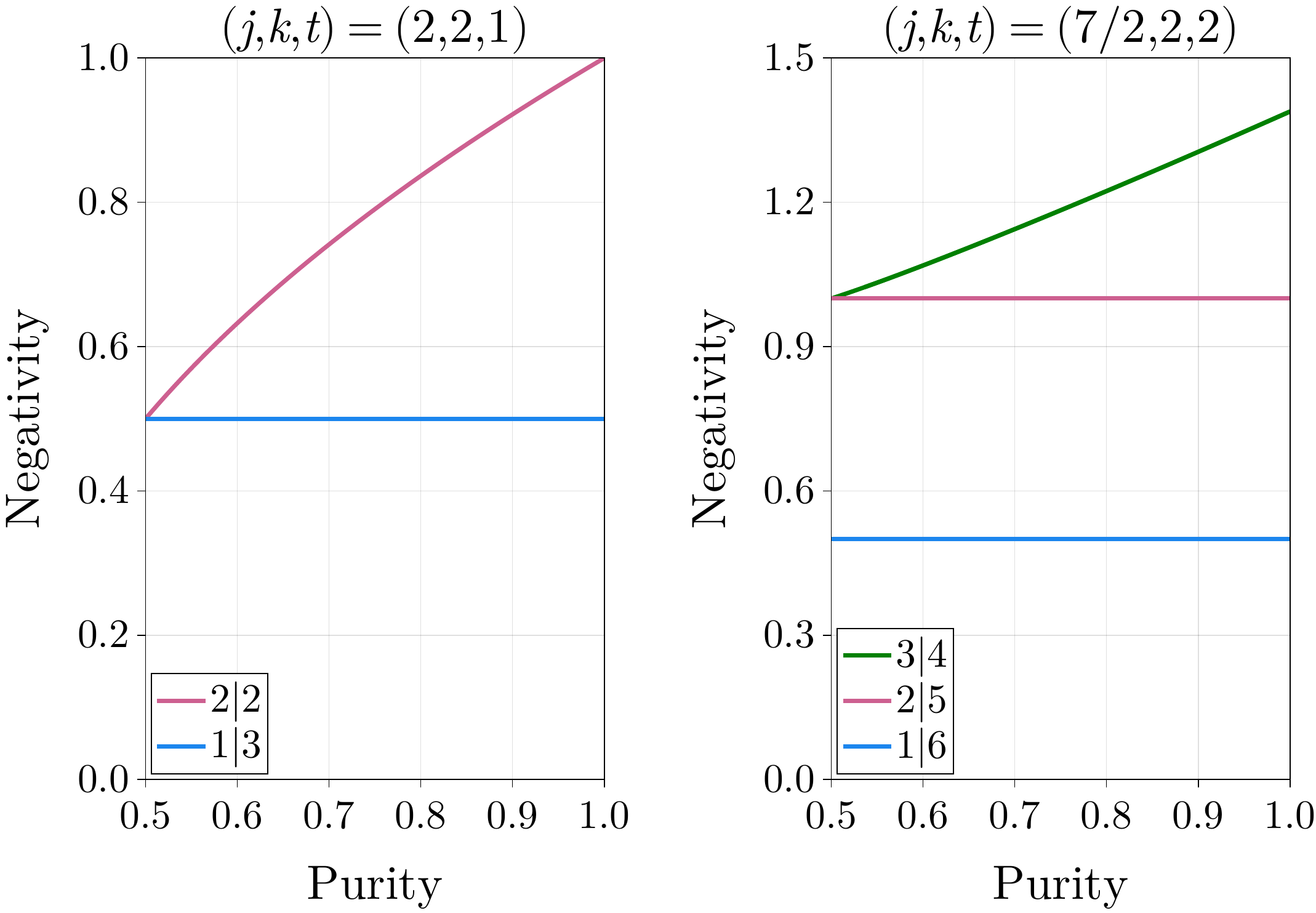}
    \caption{\label{Fig.Neg}
     Negativity of the states $\rho = \la_1 \rho_{\psi_1} + (1-\la_1) \rho_{\psi_2}$, constructed by mixing the basis states of (left panel) the $(j,k,t)=(2,2,1)$ AC subspace given in Eq.~\eqref{j2k2t1.1stplane}, and (right panel) the $(j,k,t)=(7/2,2,2)$ AC subspace given in Eq.~\eqref{Ex.j7h222.subspace}, as a function of their purity. 
    }
\end{figure}
An important consequence of (the proof of) Theorem~\ref{Theorem.1} is an upper bound on the dimension of $(j,k,t)$ AC subspaces, given by
\begin{equation}
\label{Eq.Upper.bound.kmax}
    k_{\max} \leqslant \left\lfloor \frac{2j-t+1}{t+1} \right\rfloor \, .
\end{equation}
In particular,  for $1$-AC subspaces, it gives a stricter bound $k_{\max} \leqslant \lfloor j \rfloor$ than the one given in Corollary 2 of Ref.~\cite{Per.Pau:17}. We plot this upper bound in Fig.~\ref{fig.kmax.subspaces}. On the other hand, we plot the negativity for some $(j,k,t)$ AC subspaces presented in the previous section in Fig.~\ref{Fig.Neg}. We observe the constant value of the negativity for the bipartitions $t|N-t$ associated to the anticoherence order of the AC subspaces but not for the others which decrease when the purity decreases. In summary, mixed OQRs, which come from $1$-AC subspaces, are maximally entangled states with respect to the negativity for the bipartition $1|N-1$, despite their variable purity.
\section{Conclusions}
\label{Sec.Conclusions}
We have established the necessary conditions, summarized in Table~\ref{tab.OQR}, for a mixed spin-$j$ state to be an optimal sensor of infinitesimal rotations, as prescribed by the \rema{fidelity or by the} QCRB averaged over all rotation axes~\eqref{Eq.QIF.ave}-\eqref{QCRB.Eq}. \rema{The mixed-state scenario brings the additional condition~\eqref{Cond.1} that $\IMA(\rho)$ must be a $1$-AC subspace, thus elevating $1$-AC subspaces to the status of essential ingredient in mixed-state rotation sensing.}
This has motivated our study of the existence of $t$-AC subspaces and, in particular, the maximum dimension that $1$-AC subspaces can have for a given value of $j$ \rema{(which is displayed in Fig.~\ref{fig.kmax.subspaces}, top panel)}. This value is important in the light of our results because it sets a lower bound on the purity of mixed OQRs. Our detailed study of mixed OQRs for small spins shows that the first mixed OQR is found for spin $j=2$ (see Subsec.~\ref{sec:j2}). Furthermore, we have presented various examples of AC subspaces and mixed OQRs, including $2$-AC subspaces that generate a continuous family of mixed OQRs, and cases where the mixed OQRs arise from a mixture of non-OQR pure states (see Eq.~\eqref{Eq.331.mixedOQR}). We have shown how to construct 1- and $2$-AC subspaces of any dimension for sufficiently high spin by providing explicit bases for them (see Tables~\ref{tab:1AC.subspaces}--\ref{tab:2AC.subspaces} for selected examples). We have proved that the dimension of these subspaces increases linearly with the spin quantum number $j$, and we have compared it with the maximum dimension $k_{\max}$ obtained numerically for the same spin (see  Fig.~\ref{fig.kmax.subspaces}, bottom panel). In addition, we have provided the upper bound~\eqref{Eq.Upper.bound.kmax} on $k_{\max}$ as a function of the spin $j$ and the anticoherence order $t$. \rema{The existence of AC subspaces for large spin quantum numbers implies that mixed OQRs can be prepared in high-dimensional quantum systems, such as atomic many-body or multiphoton systems. Although all of our examples concern single-spin systems, they can in principle be implemented in any finite-dimensional system (e.g., multiqudit systems), as explained in Subsection~\ref{SubSec.OQR.general}.} One of the main conclusions of our work is that mixed states can achieve the same sensitivity as optimal pure states. This property can be related to entanglement by interpreting spin states as symmetric multiqubit states. More precisely, our Theorem~\ref{Theorem.1} establishes that any state obtained by mixing pure states from a $t$-AC subspace, in particular mixed OQRs, maximizes all negativities for the bipartitions $t'|N-t'$ with $t'\leqslant t$. Therefore, mixed spin-$j$ OQRs not only have genuine multipartite entanglement, as was already known to be necessary for pure states to be OQRs~\cite{PhysRevLett.102.100401,PhysRevA.85.022321,PhysRevA.85.022322}, but they are also maximally entangled states with respect to the negativity for the bipartition $1|N-1$ among all symmetric states.

\rema{In this work, we have focused on the search for optimal states for metrological tasks aimed at measuring rotations with respect to fidelity or the QCRB. OQR with respect to fidelity, $\OQRI$, is conceived in relation to a fidelity measurement protocol~\cite{Liang_2019}. On the other hand, for an OQR with respect to the QCRB, $\OQRInv$, one has to find a measurement protocol capable of achieving the ultimate sensitivity. An optimal choice of measurements for the estimation of a single parameter, whether from pure or mixed states, can be found by using the expression of the QFI in terms of the symmetric logarithmic derivative~\cite{Bra.Cav:94,RevModPhys.90.035005}. The saturation of the QCRB has been studied for pure spin-like two-mode photonic states~\cite{Z_Goldberg_2021,PhysRevApplied.20.024052}, where a projection-valued measure for estimating the axis and angle of a rotation has been found. An interesting follow-up to this work would be the implementation of an optimal measurement scheme for mixed spin states.
}
\section*{Acknowledgments}
ESE acknowledges support from the postdoctoral fellowship of the IPD-STEMA program of the University of Liège (Belgium). JM acknowledges the FWO and the F.R.S.-FNRS for their funding as part of the Excellence of Science programme (EOS project 40007526). CC would like to acknowledge partial financial support from the
DGAPA-UNAM project IN112224.
\begin{appendix}
\section{Existence of mixed anticoherent states to any order}
\label{App.t.anticoherence}
We prove here the existence of mixed AC states of any order $t$ other than the maximally mixed state. More generally, we prove the existence of mixed states with $\Tr(\rho\, T_{\si \mu})=0$ for any subset of indices $\si \in \mathcal{I} \subset \{ 1 , \dots , 2j \}$ and $\mu = -\si , \dots , \si$. We start with the maximally mixed state $\rho_0 =\mathds{1}/(2j+1)$ and a mixture  
\begin{equation}
    \rho = \rho_0 + \epsilon A \, ,
\end{equation}
with an operator $A$ such that
\begin{equation}
    A= \sum_{\substack{L=1\\[0.1cm] L\notin \mathcal{I}} }^{2j} \sum_{M=-L}^L A_{LM} T_{LM} \, ,
\end{equation}
with arbitrary coefficients $A_{LM}$ making $A$ Hermitian. The state $\rho$ is Hermitian, with trace $\Tr (\rho)=1$, and positive for a sufficiently small $\epsilon >0$. In fact, if we denote by $\la_{\min}$ the lowest (and therefore potentially negative) eigenvalue of $A$, then $\rho$ is a valid state for $\epsilon \leq [(2j+1)|\la_{\min}|]^{-1}$. In particular, considering $\mathcal{I}= \{ 1, 2, \dots ,t\}$, the state defined above is $t$-AC.
\section{2-AC spin-$3/2$ mixed states}
\label{2Ant.3h2spin}
Let us characterize the set of $2$-AC spin-$3/2$ mixed states. By definition, they have an expansion in the multipole operator basis of the form
\begin{equation}
    \rho = \frac{1}{4}\mathds{1} + \sum_{M= -3}^3 \rho_{3 M} T_{3 M } \, .
\end{equation}
We can reduce the number of $\rho_{3M}$ variables by a global rotation rendering $\rho_{33}=0$ and $\rho_{32} \in \mathds{R}$. Moreover, by the hermiticity of the state, we also get $\rho_{3,-3}=0$ and $\rho_{32} \in \mathds{R}$. We now perform the change of variables 
\begin{equation}
\begin{aligned}    
&   \rho_{30}=w\, c_0 \, , \quad
    \rho_{31}= -\rho^*_{3-1} = w\, c_1 e^{i \phi} 
    \, ,
\\
&    \rho_{32}= \rho_{3-2}= w\, c_2 \, , 
\end{aligned}
\end{equation}
so that
\begin{multline}     
    \rho = \frac{\mathds{1}}{4} + w \Big[c_0 T_{30} + c_1  \left( e^{i \phi} T_{31} - e^{-i \phi} T_{3-1} \right)
    \\
    + c_2 \left( T_{32}+ T_{3-2} \right) \Big] \, ,
\end{multline} 
with $c_0^2 + 2c_1^2 + 2c_2^2 =1$ and where all the new variables $(w,c_0,c_1,c_2,\phi)$ are real. Finally, the  eigenvalues of $\rho$ 
 \begin{equation}
     \begin{aligned}
         \la_{\pm \pm} = \frac{1}{4} \pm \frac{w}{2} \sqrt{1 \pm \frac{2}{5}\sqrt{X}} 
     \end{aligned}
 \end{equation}
 must be non-negative, where
 \begin{multline}         
     X= 2 c_1^2 \left(-2 \sqrt{30} c_2 c_0 \cos (2 \phi )+8 c_0^2+15 c_2^2\right)
     \\ 
     + 21 c_1^4+4 c_0^2 \left(c_0^2+10 c_2^2\right).
\end{multline}
This is guaranteed if the lowest eigenvalue $\la_{-+} $ is non-negative, which is fulfilled for 
$w \leqslant \sqrt{5}/ 2 \sqrt{5+ 2 \sqrt{X}}$.
In particular, if we are interested in the $2$-AC state with the highest purity $\Tr(\rho^2)= \frac{1}{4} + w^2$, we must
maximize the variable $w$ for valid states, i.e., for states with $\la_{-+} \geqslant 0$. To fulfill this, we minimize the factor of the variable $w$ in $\la_{-+}$, $\sqrt{1+ \frac{2}{5}\sqrt{X}}$, achieved for $X=0$. The minimum is given by $(c_0 , c_1 , c_2 , \phi) = (0,0,\frac{1}{\sqrt{2}},0)$ and the corresponding state has doubly-degenerated eigenvalues $\la_{\pm \pm}= \frac{1}{4} \pm \frac{w}{2}$. Then, we can take $w=\frac{1}{2}$, leading to the state
\begin{equation}
    \rho = \frac{1}{2} \left( \ket{\psi_1} \bra{\psi_1} + \ket{\psi_2} \bra{\psi_2} \right) \, ,
\end{equation}
with eigenvectors~\eqref{St.s3h2} and purity $\Tr(\rho^2) = \frac{1}{2}$.
\section{Uniqueness of the \texorpdfstring{$(2,2,1)$}{Lg} AC subspace}
\label{App.uniqueness.spin2}
In this Appendix, we prove that, up to a rotation, the $(2,2,1)$ AC subspace $V$ spanned by the states~\eqref{j2k2t1.1stplane} is unique. We begin by considering a generic $(2,2,1)$ AC subspace $V$ with basis states $\{ \ket{\psi_1} , \ket{\psi_2} \}$. By definition, the $\ket{\psi_k} $ states are $1$-AC. We then rotate the subspace to $\tilde{V}$, now with basis states $\ket{\tilde{\psi}_k} = \Rr \ket{\psi_k}$, such that $\ket{\tilde{\psi}_1} $ has the form of a generic spin-$2$ $1$-AC state (as given in Ref.~\cite{Bag.Dam.Gir.Mar:15})
\begin{equation}
    \ket{\tilde{\psi}_1} = \frac{1}{\sqrt{2+|\mu|^2}}(1,0,\mu , 0,1) \, ,
\end{equation}
where $\mu \in \mathds{C}$. The unnormalized vectors $J_{a}\ket{\tilde{\psi}_1}$ for $a=x,y,z$ read
\begin{equation}
\begin{aligned}
  J_x \ket{\tilde{\psi}_1} \propto & \left( 1 + \sqrt{\tfrac{3}{2}}\, \mu \right) (0,1,0,1,0) \, ,
\\    
  J_y \ket{\tilde{\psi}_1} \propto & \left( 1 - \sqrt{\tfrac{3}{2}}\, \mu \right) (0,1,0,-1,0) \, ,
\\
  J_z \ket{\tilde{\psi}_1} \propto &{} (1,0,0,0,-1) .
\end{aligned}
\end{equation}
We  see that for $\mu \neq \pm \sqrt{\frac{2}{3}}$, the set
\begin{equation}
B_{\tilde{V}} = \left\{ \ket{\tilde{\psi}_1} , J_x\ket{\tilde{\psi}_1} , J_y\ket{\tilde{\psi}_1} , J_z\ket{\tilde{\psi}_1} \right\}  
\end{equation}
consists of 4 orthogonal vectors. Its orthogonal complement $B_{\tilde{V}}^{\perp}$ has dimension $1$ and, by direct calculation, can be seen to be generated by the state
\begin{equation}
\frac{1}{\sqrt{4+2 |\mu|^2}}\left( \mu^{*},0,-2 ,0 , \mu^{*} \right)
\, .
\end{equation}
The latter state must be equal to $\ket{\tilde{\psi}_2}= \Rr \ket{\psi_2}$ because $\tilde{V}$ is a $1$-AC subspace and must obey the condition~\eqref{Eq.Cond.1AC.subspace}. Explicit calculation of the states $\ket{\tilde{\psi}_k}$ shows that $\tilde{V}$ is the same subspace as that spanned by the states~\eqref{j2k2t1.1stplane} whatever the value of $\mu$. In the special case $\mu = \pm \sqrt{2/3}$, these states are, up to a rotation, equal to the $\ket{2,0}$ state~\cite{Bag.Dam.Gir.Mar:15}. We can then study this case by starting with $\ket{\tilde{\psi}_1}= (0,0,1,0,0)$. Generating the set $B_{\tilde{V}}$ in the same way, which now contains 3 vectors, we obtain that $\ket{\tilde{\psi}_2} \in B_{\tilde{V}}^{\perp}$ must have the form
 \begin{equation}
     \ket{\tilde{\psi}_2} \propto (a,0,0,0,b) \, ,
 \end{equation}
with $a,b \in \mathds{C}$. However, since $     \ket{\tilde{\psi}_2}$ must be $1$-AC, we must have that $b=a e^{i \phi}$. Consequently, $\tilde{V}$ is spanned by the states
\begin{equation}
    (0,0,1,0,0) \, , \quad \frac{1}{\sqrt{2}} \left(e^{-i \phi/2},0,0,0,e^{i \phi/2} \right) \, ,
\end{equation}
which, after a rotation by $-\phi/2$ about the $z$ axis, is the subspace spanned by the states
\begin{equation}
    (0,0,1,0,0) \, , \quad \frac{1}{\sqrt{2}} \left(1,0,0,0,1 \right) \, .
\end{equation}
Again, this AC subspace is equivalent to the subspace spanned by the states~\eqref{j2k2t1.1stplane}.
\section{Examples of 1-AC subspaces}
\label{Appendix1ACsub}
\subsection{\texorpdfstring{$(5/2,2,1)$}{Lg} AC subspaces}
\rema{There are several inequivalent $(5/2,2,1)$ AC subspaces. A first example is the space $V_1$ spanned by the states
\begin{equation}
\label{Ex1.N5k2t1}
\begin{aligned}    
    \ket{\psi_1}={}& \frac{1}{\sqrt{8}} \left( \sqrt{3},0,0,0,\sqrt{5},0 \right) \, , 
    \\
    \ket{\psi_2}={}& \frac{1}{\sqrt{8}} \left( 0,-\sqrt{5},0,0,0,\sqrt{3} \right) \, .
\end{aligned}
\end{equation}
The candidates for mixed OQRs are given by a mixture of states forming a generic basis of $V_1$, 
\begin{equation}
\rho = \la_1 \ket{\tilde{\psi}_1}\bra{\tilde{\psi}_1}+\la_2\ket{\tilde{\psi}_2}\bra{\tilde{\psi}_2}
\end{equation}
with $\la_1+\la_2=1$ and
\begin{equation}
\begin{aligned}
\ket{\tilde{\psi}_1} = &
\cos \left( \tfrac{\theta}{2} \right) \ket{\psi_1} + 
\sin \left( \tfrac{\theta}{2} \right) e^{i \phi} \ket{\psi_2} , 
\\[3pt]
\ket{\tilde{\psi}_2} = &    
\sin \left( \tfrac{\theta}{2} \right) \ket{\psi_1} -
\cos \left( \tfrac{\theta}{2} \right) e^{i \phi} \ket{\psi_2},
\end{aligned}
\end{equation}
where $\theta$ and $\phi$ are real parameters. The function $\mathcal{A}_2(\rho)$ for these mixtures is calculated to be
\begin{equation}
\mathcal{A}_2(\rho) = 
\frac{9}{128} \left[ 
\left(\la_1 - \lambda_2\right){}^2 \cos (2 \theta ) + 4 \lambda_1 \lambda_2 +13
\right].
\end{equation}
Its maximum value $\mathcal{A}_2 = 63/64 \approx 0.984375$ is reached for $\la_1 = \la_2 = 1/2$ and any $\theta , \phi$ values, or for $\theta=0$ and any weights $\la_1 , \la_2$ and phase $\phi$. 
The corresponding mixed states are almost $2$-AC, but never fulfill condition~\eqref{Cond.2}. Hence, they are not mixed $\OQRInv$.}

\rema{A second example of a $(5/2,2,1)$ AC subspace is the vector space $V_2$ spanned by the states
\begin{equation}
\label{Ex2.N5k2t1}
\begin{aligned}
    \ket{\psi_1} ={}&  \frac{1}{4} \left( 
    0, \alpha,0,- \beta,0, \gamma
    \right) \, ,
\\
    \ket{\psi_2} ={}&
\frac{1}{4}
\left( 
 \gamma,0, \beta,0, \alpha,0
\right).
\end{aligned}
\end{equation}
with $\alpha=\sqrt{20-5 \sqrt{7}}$, $\beta=\sqrt{10\sqrt{7}-20}$ and $\gamma=\sqrt{16-5 \sqrt{7}}$. Again, we search for a $2$-AC state $\rho$ in the form of a convex combination of basis states of a generic basis of $V_2$. We find that the maximum of $\mathcal{A}_2 (\rho)$ is equal to
\begin{equation}
\mathcal{A}_2 =     
\frac{3}{32} \left(7 \sqrt{7}-8\right) \approx 0.9863 \, ,
\end{equation}
achieved for the equally weighted mixture ($\la_1=\la_2=1/2$), irrespective of the basis of $V_2$. Once again, $\max\mathcal{A}_2(\rho)\ne 1$ and the corresponding state is not $2$-AC and is therefore not a mixed OQR.}

\rema{Although there are mixed $2$-AC states (see Appendix~\ref{App.t.anticoherence}), we observe numerically that none of them has an image $\IMA(\rho)$ forming a $1$-AC subspace and, consequently, we have strong evidence that there are no OQRs of spin $5/2$ with respect to the QCRB (see also Fig.~\ref{fig.kmax.subspaces}).}
\subsection{\texorpdfstring{$(3,3,1)$}{Lg} AC subspace}
\rema{The first $3$-dimensional $1$-AC subspace $V$ appears for $j=3$, and is spanned by the states
\begin{equation}
\label{j3k3t1.Ex1}
\begin{aligned}
    & \ket{\psi_1} = \sqrt{\tfrac{2}{5}}\, \left| 3,3\right\rangle +
    \sqrt{\tfrac{3}{5}}\, \left| 3,-2 \right\rangle \, , \\
    & \ket{\psi_2} = -\sqrt{\tfrac{3}{5}}\, \left| 3,2 \right\rangle +
    \sqrt{\tfrac{2}{5}}\, \left| 3,-3 \right\rangle \, , \\
    & \ket{\psi_3} = \left| 3,0\right\rangle \, .\phantom{\sqrt{\tfrac{2}{5}}}
\end{aligned}
\end{equation}
The measure of $2$-anticoherence $\mathcal{A}_2$ of a general coherent superposition $|\psi\rangle=a|\psi_1\rangle+b|\psi_2\rangle+c|\psi_3\rangle$, with  $a,b,c\in\mathbb{C}$ and $|a|^2+|b|^2+|c|^2=1$, is given by
\begin{equation}
\label{A2psiabc}
\mathcal{A}_2(\rho_{\psi})=\frac{3}{25} \left[8-4 | a| ^2 | b| ^2-| c| ^4+4\, \mathrm{Re}\big(a b \left(c^*\right)^2\big)\right] \, .
\end{equation}
The maximum of $\mathcal{A}_2(\rho_{\psi})$ under the normalization constraint is equal to $24/25$~\footnote{The maximum is achieved, e.g., for $a=b=1/2$ and $c=1/\sqrt{2}$.}. Hence, no pure state $|\psi\rangle\in V$ is $2$-AC. On the other hand, the statistical mixture of the states given in Eq.~\eqref{j3k3t1.Ex1} with weights $\la_k$ yields a $\rho$ with $\mathcal{A}_2(\rho)= 3(2-\la_3)(4+3\la_3)/25$. This then shows that for $\la_3=1/3$, $\rho$ is $2$-AC and therefore a mixed OQR. This family of $\OQRInv$, explicitly given by
\begin{equation}
\label{Eq.331.mixedOQR}
\rho = \la_1 \rho_{\psi_1} + \left(\frac{2}{3}-\la_1 \right) \rho_{\psi_2} +
\frac{1}{3} \rho_{\psi_3},
\end{equation}
with $\lambda_1\in [0,2/3]$, is the first example of mixed OQRs without any pure OQR in their image $\IMA( \rho)$.}
\subsection{\texorpdfstring{$(9/2,4,1)$}{Lg} AC subspace}
\rema{The first $4$-dimensional $1$-AC subspace $V$ appears for $j=9/2$, and is spanned by the states
\begin{equation}
\label{1AC4planej9o2}
    \begin{aligned}
        & \ket{\psi_1} = \left(0,0,0,\tfrac{\sqrt{\frac{5}{2}}}{2},0,0,0,-\tfrac{\sqrt{\frac{3}{2}}}{2},0,0\right),\\
        & \ket{\psi_2} = \left(0,0,-\tfrac{\sqrt{\frac{3}{2}}}{2},0,0,0,-\tfrac{\sqrt{\frac{5}{2}}}{2},0,0,0\right),\\
        & \ket{\psi_3} = \left(\tfrac{1}{4} \sqrt{\tfrac{11}{2}}\, e^{i \zeta},0,0,0,\tfrac{\sqrt{3}}{4},0,0,0,\tfrac{\sqrt{\frac{15}{2}}}{4},0\right),\\
        & \ket{\psi_4} = \left(0,\tfrac{\sqrt{\frac{15}{2}}}{4},0,0,0,-\tfrac{\sqrt{3}}{4},0,0,0,\tfrac{1}{4} \sqrt{\tfrac{11}{2}} \, e^{i \zeta}\right)
    \end{aligned}
\end{equation}
with $\zeta=\tan ^{-1}\left(2 \sqrt{\frac{7}{5}}\right)$.}
\section{Examples of 2-AC subspaces}
\rema{In this subsection, we give some examples of $2$-AC subspaces for small $j$ values.}
\subsection{\texorpdfstring{$(7/2,2,2)$}{Lg} AC subspace}
\rema{The first instance of a $2$-AC subspace is found for $j=7/2$. An example is the space spanned by the states
\begin{equation}
\label{Ex.j7h222.subspace}
\begin{aligned}    
    \ket{\psi_1} ={}& \sqrt{\tfrac{3}{10}}\, \left| \tfrac{7}{2},\tfrac{7}{2} \right\rangle +
    \sqrt{\tfrac{7}{10}}\, \left| \tfrac{7}{2},-\tfrac{3}{2} \right\rangle  , 
    \\
    \ket{\psi_2} ={}& \sqrt{\tfrac{7}{10}}\, \left| \tfrac{7}{2},\tfrac{3}{2} \right\rangle -
    \sqrt{\tfrac{3}{10}}\, \left| \tfrac{7}{2},-\tfrac{7}{2} \right\rangle  .
\end{aligned}
\end{equation}}
\subsection{\texorpdfstring{$(5,2,2)$}{Lg} AC subspace}
\rema{For spins larger than $j=7/2$, we did not find numerically other $(j,2,2)$ AC subspaces up to the value $j=5$. One example is the space spanned by the states
\begin{equation}
\label{Ex.522.subspace}
\begin{aligned}    
    \ket{\psi_1}={}&  
    \sqrt{\tfrac{2}{7}}\, \left| 5 , 5 \right\rangle + \sqrt{\tfrac{5}{7}}\, \left| 5 , -2 \right\rangle 
   , 
    \\ 
\ket{\psi_2} ={}&  
    -\sqrt{\tfrac{5}{7}}\, \left| 5 , 2 \right\rangle + \sqrt{\tfrac{2}{7}}\, \left| 5 , -5 \right\rangle .
\end{aligned}
\end{equation}}
\subsection{\texorpdfstring{$(7,3,2)$}{Lg} AC subspace}
\rema{A $2$-AC subspace of dimension $3$ is first found for spin $7$ and is spanned by the states
\begin{equation}
\label{Ex.j732.subspace}
    \begin{aligned}
        & \ket{\psi_1} = (0,0,a,0,0,b,0,0,c,0,0,d,0,0,e) , \\
        & \ket{\psi_2} = (e,0,0,d,0,0,c,0,0,b,0,0,a,0,0) , \\
        & \ket{\psi_3} = (0,f,0,0,g,0,0,h,0,0,g,0,0,f,0) ,
    \end{aligned}
\end{equation}
with
\begin{equation}
\begin{aligned}
& a \approx -0.4604769924899385+0.2090691916016556 i,\\
    & b \approx 0.2215035777892046-0.4925631870366248 i,\\
    & c \approx 0.3036273245094665+0.3014334601129537 i,\\
    & d \approx -0.1069092203207547+0.2403277996106323 i,\\
    & e \approx -0.3180190622270446-0.3149505431871214 i,\\
    & f \approx 0.4395729087324888+0.1474365706612073 i,\\
    & g \approx 0.4058433985142867+0.1117167627487395 i,\\
    & h \approx 0.4038887232746600+0.2292839547339512 i .
    \end{aligned}
\end{equation}}
\section{Maximal negativity of mixed states}
\label{App.Neg.Sub}
Before proving Theorem~\ref{Theorem.1}, let us examine the properties of an orthonormal basis $B_V= \{ \ket{\psi_i} \}_{i=1}^{k}$ associated with a $(j,k,t)$ AC subspace $V$ with $t\leqslant j$. Since each state $\ket{\psi_i}$ is $t$-AC, its Schmidt decomposition for the bipartition $A|B\equiv t|N-t$ 
is given by (see Subsection~\ref{SubSec.Neg})
    \begin{equation}
    \label{Schmidt.dec.mix}
        \ket{\psi_m} = \frac{1}{\sqrt{t+1}} \sum_{\alpha=1}^{t+1} \ket{\phi^{A}_{\alpha}} \ket{\phi^{B}_{m\alpha}} \, ,
    \end{equation}
    where we are free to choose the common basis $\{ \ket{\phi^{A}_{\alpha}} \}$ of $\Hs^{A}$ for each state $\ket{\psi_m}$, using the non-uniqueness of the singular value decomposition on the degenerated Schmidt eigenvalues~\cite{hor.joh:12}.
  Here, all the Schmidt numbers are degenerated and, consequently, we can choose the basis of $\Hs^{A}$ for each state $\ket{\psi_m}$. We now establish a consequence of the anticoherence property of a subspace $V$ over the states $\ket{\phi^{B}_{m\alpha}}$ in the decomposition~\eqref{Schmidt.dec.mix}.
\begin{lemma}
\label{Lemma.Schmidt}
    Let $V$ be a $(j,k,t)$ AC subspace, with an orthonormal basis $B_V = \{ \ket{\psi_m} \}_{m=1}^{k}$ where each state has a Schmidt decomposition in the bipartition $t|N-t$, $\Hs^{(j)} \subset \Hs^{A} \otimes \Hs^{B} = \Hs^{(\frac{t}{2})} \otimes \Hs^{(j-\frac{t}{2})} $, given by
    \begin{equation}
        \ket{\psi_m} = \frac{1}{\sqrt{t+1}} \sum_{\alpha=1}^{t+1} \ket{\phi^{A}_{\alpha}} \ket{\phi^{B}_{m\alpha}} \, .
    \end{equation}
Then, the set $\{ \ket{\phi_{m\alpha}^B}\}_{m,\alpha}$ is orthonormal
\begin{equation}
    \left\langle \phi_{m \alpha}^{B} |
    \phi_{n \beta}^{B}
    \right\rangle = \delta_{mn} \delta_{\alpha \beta} \, .
\end{equation}
\end{lemma}
\emph{Proof.} The Schmidt decomposition of each state $\ket{\psi_m}$ implies that each subset $\{ \ket{\phi_{m\alpha}^B} \}_{\alpha =1}^{t+1}$ is orthonormal $\langle \phi_{m\alpha}^B | \phi_{m\beta}^B \rangle = \delta_{\alpha \beta}$. Now, due to the fact that $V$ is a $(j,k,t)$ AC subspace, each state defined by a linear combination of two states of $B_V$,
\begin{equation}
    c_1 \ket{\psi_m} + c_2 \ket{\psi_n}
= \frac{1}{\sqrt{t+1}} \sum_{\alpha = 1}^{t+1} \ket{\phi_{\alpha}^{A}} \left( c_1 \ket{\phi_{m\alpha}^B} + c_2 \ket{\phi_{n\alpha}^B} \right) \, ,
\end{equation}
is $t$-AC, and then, it has a fully degenerated Schmidt decomposition. This implies that the states $c_1 \ket{\phi_{m\alpha}^B} + c_2 \ket{\phi_{n\alpha}^B}$ must be orthogonal,
\begin{equation}
\left( c_1^* \bra{\phi_{m\beta}^B} + c_2^* \bra{\phi_{n\beta}^B} \right)
    \left( c_1 \ket{\phi_{m\alpha}^B} + c_2 \ket{\phi_{n\alpha}^B} \right) = \delta_{\alpha \beta},
\end{equation}
for any values of $c_1$ and $c_2$. In particular, we can choose the following two pairs $c_1 = 1/\sqrt{2}$, $c_2 = 1/\sqrt{2}$ and $c_1 = 1/\sqrt{2}$, $c_2 = i/\sqrt{2}$ to arrive at
\begin{equation}
    \langle \phi_{n\beta}^B | \phi_{m\alpha}^B \rangle = \pm
    \langle \phi_{m \beta}^B | \phi_{n \alpha}^B \rangle \, .
\end{equation}
Consequently, $\langle \phi_{n\beta}^B | \phi_{m\alpha}^B \rangle =0$ for any $m , n$ which concludes the proof. 
$\square$

As an example, take the $(2,2,1)$ AC subspace spanned by the states
\begin{equation}
    \begin{aligned}
        \ket{\psi_1} &= \frac{1}{\sqrt{3}} \left( \ket{2,2} + \sqrt{2} \ket{2,-1} \right) \, ,
        \\
        \ket{\psi_2} &= \frac{1}{\sqrt{3}} \left(\sqrt{2}\ket{2,1} -\ket{2,-2} \right) \, ,
    \end{aligned}
\end{equation}
which is equivalent to the subspace defined in Eq.~\eqref{j2k2t1.1stplane} up to a global rotation. Without loss of generality, we can take the basis for the Schmidt decomposition in the partition $1|3$ of the subsystem $\Hs^{A}$ as $\left| \phi_{m}^A \right\rangle = \left| \frac{1}{2} , \pm \frac{1}{2} \right\rangle $. The Schmidt decomposition of the states are $\ket{\psi_m} =  \left( \ket{\phi_+^A} \ket{\phi_{m+}^B} + \ket{\phi_-^A} \ket{\phi_{m-}^B}  \right) / \sqrt{2}$ with
\begin{equation*}
    \ket{\phi_{1+}^B} =  \frac{\sqrt{2} \left| \frac{3}{2} \frac{3}{2} \right\rangle 
    + \left|\frac{3}{2},-\frac{3}{2} \right\rangle }{\sqrt{3}} 
    , \quad
    \ket{\phi_{1-}^B} = \left| \tfrac{3}{2} , - \tfrac{1}{2} \right\rangle
    \, ,
\end{equation*}
\begin{equation}
\ket{\phi_{2+}^B} =  
     \left| \tfrac{3}{2} ,  \tfrac{1}{2} \right\rangle
     , \quad
    \ket{\phi_{2-}^B} =  
 \frac{ \left|\frac{3}{2},\frac{3}{2} \right\rangle - \sqrt{2} \left| \frac{3}{2} -\frac{3}{2} \right\rangle  }{\sqrt{3}} \, , 
\end{equation}
where we observe that the four spin-$3/2$ states $\{ \ket{\phi_{m\alpha}^B}\}_{m,\alpha}$ are indeed orthonormal.

By Lemma~\ref{Lemma.Schmidt}, the existence of a $(j,k,t)$ AC subspace implies the existence of an orthonormal set of $k(t+1)$ states in $\Hs^B =\Hs^{(j-\frac{t}{2})}$. Then, the dimension of $\Hs^{(j-\frac{t}{2})}$ implies the inequality $k(t+1) \leqslant \dim (\Hs^B) = 2j-t+1$. We can use this inequality to obtain the upper bound on the maximal dimension of a $(j,k,t)$ AC subspace mentioned in Eq.~\eqref{Eq.Upper.bound.kmax}.

Now let us prove Theorem~\ref{Theorem.1} following an argument similar to that used for pure states in~\cite{JOHNSTON20181,Den.Mar:22}.

\emph{Proof of Theorem~\ref{Theorem.1}}. Let us consider a mixed state $\rho= \sum_{m=1}^k \lambda_m \ket{\psi_{k}} \bra{\psi_k} $ where $\IMA(\rho)$ is a $(j,k,t)$ AC subspace $V$. The state $\rho$ seen as a symmetric $N$-qubit state of a bipartite system $t|N-t$ reads
\begin{equation}
    \rho = \frac{1}{t+1} \sum_{m=1}^k \sum_{\alpha ,\beta=1}^{t+1} \lambda_m \ket{\phi_{\alpha}^A} \ket{\phi_{m\alpha}^B} \bra{\phi_{\beta}^A} \bra{\phi_{m\beta}^B} \, ,
\end{equation}
where $\{ \ket{\phi_{\alpha}^A}\}_{\alpha}$ is an orthonormal basis of $\Hs^A$ by definition of the Schmidt decomposition, and $C = \{ \ket{\phi_{m \alpha}^B}\}_{m,\alpha}$ is an orthonormal set (from Lemma~\ref{Lemma.Schmidt}) with $k(t+1)$ elements. The partial transpose state $\rho^{T_B}$ reads
\begin{equation}
    \rho^{T_B} = \frac{1}{t+1} \sum_{m=1}^k \sum_{\alpha ,\beta=1}^{t+1} \lambda_m \ket{\phi_{\alpha}^A} \ket{\phi_{m\beta}^{B*}} \bra{\phi_{\beta}^A} \bra{\phi_{m\alpha}^{B*}} \, ,
\end{equation}
where the asterisk means complex conjugation. The $(t+1)(2j-t+1)$ eigenvectors of $\rho^{T_B}$ are divided in four types:
\begin{enumerate}
    \item[$\bullet$] $(t+1)k$ eigenvectors $\ket{x_{m\alpha}} \equiv \ket{\phi^A_{\alpha}} \ket{\phi^{B*}_{m\alpha}}$ for $\alpha= 1 , \dots , t+1$, $m=1,\dots, k$ and eigenvalues $\lambda_k / (t+1)$.
    
    \item[$\bullet$] {two sets of $t(t+1)k/2$ eigenvectors given by
    \begin{equation}
        \quad\quad\ket{y^{\pm}_{\alpha , \beta ,m}} \equiv \frac{1}{\sqrt{2}}
        \left( 
        \ket{\phi^A_{\alpha}} \ket{\phi^{B*}_{m\beta}} \pm
        \ket{\phi^A_{\beta}} \ket{\phi^{B*}_{m\alpha}}
        \right) \, ,
    \end{equation}
    for $\alpha < \beta$ with $\alpha,\beta =1 , \dots , t+1$ and $m=1,\dots , k$, with eigenvalues $\pm \lambda_m/ (t+1)$.}\\[0pt]
    \item[$\bullet$] In the case that $C^{\perp}$ is not empty, it has dimension $h= 2j-k(t+1)-t+1$ with a basis labeled by $\ket{\varphi^{B}_{m}}$ with $m=1,\dots , h $. Then there exist $h(t+1)$ eigenvectors of $\rho^{T_B}$, $\ket{z_{\alpha m}}= \ket{\phi_{\alpha}^A} \ket{\varphi^{B *}_{m}}$ with eigenvalues equal to zero.
\end{enumerate}
We note that the only negative eigenvalues come from the eigenvectors $\ket{y^-_{\alpha,\beta,m}}$. The negativity of $\rho$ is thus equal to
\begin{equation}
    \Nega_t (\rho) = \sum_{m=1}^k \sum_{\substack{\alpha,\beta=1 \\ \alpha < \beta}}^{t+1}
    \frac{\lambda_m}{t+1} = \frac{t}{2} ,
\end{equation} 
i.e., equal to the negativity~\eqref{Eq.Max.Neg.Pure} of a pure $t$-AC state, which is maximal among symmetric states. $\square$
\end{appendix}
\bibliographystyle{apsrev4-2}
\bibliography{Refs_APP}
\end{document}